%% file: conference_101719.tex
\documentclass[conference]{IEEEtran}
\IEEEoverridecommandlockouts
\usepackage{cite}
\usepackage{amsmath,amssymb,amsfonts}
\usepackage{algorithmic}
\usepackage{graphicx}
\usepackage{textcomp}
\usepackage{xcolor}
\def\BibTeX{{\rm B\kern-.05em{\sc i\kern-.025em b}\kern-.08em
    T\kern-.1667em\lower.7ex\hbox{E}\kern-.125emX}}

\usepackage{tikz}
\usepackage{stackengine}
\usepackage{caption}
\usepackage{subcaption}
\usepackage{hyperref}
\usepackage{pgfplots}
\usepackage[acronym]{glossaries}
\usepackage{booktabs}
\usepackage[ruled,vlined]{algorithm2e}

\usepackage{bm}

\input{00_Commands}

\input{00_Glossary}
    
\begin{document}

\title{On the Design of Diffusion-based\\ Neural Speech Codecs
%{\footnotesize \textsuperscript{*}Note: Sub-titles are %not captured in Xplore and
%should not be used}
%\thanks{Identify applicable funding agency here. If none, %delete this.}
}

\author{

\IEEEauthorblockN{Pietro Foti and Andreas Brendel}
\IEEEauthorblockA{Fraunhofer IIS
Erlangen (pietro.foti@iis.fraunhofer.de)}

%\and

%\IEEEauthorblockN{Andreas Brendel}
%\IEEEauthorblockA{\textit{Fraunhofer IIS} \\
%Erlangen, Germany}

\thanks{This work has been supported by the Free State of Bavaria in the DSgenAI project.
}

}

\maketitle

\begin{abstract}

Recently, neural speech codecs (NSCs) trained as generative models have shown superior performance compared to conventional codecs at low bitrates.
Although most state-of-the-art NSCs are trained as Generative Adversarial Networks (GANs), Diffusion Models (DMs), a recent class of generative models, represent a promising alternative due to their superior performance in image generation relative to GANs. Consequently, DMs have been successfully applied for audio and speech coding among various other audio generation applications. However, the design of diffusion-based NSCs has not yet been explored in a systematic way. We address this by providing a comprehensive analysis of diffusion-based NSCs divided into three contributions. First, we propose a categorization based on the conditioning and output domains of the DM. This simple conceptual framework allows us to define a design space for diffusion-based NSCs and to assign a category to existing approaches in the literature. Second, we systematically investigate unexplored designs by creating and evaluating new diffusion-based NSCs within the conceptual framework. Finally, we compare the proposed models to existing GAN and DM baselines through objective metrics and subjective listening tests.

\end{abstract}


\section{Introduction}

\acrlong{nsc}s (\acrshort{nsc}s) have been significantly advanced in recent years, offering improved audio quality and compression efficiency compared to traditional codecs, especially for low and very low bitrates. Most \acrfull{sota} \acrshort{nsc}s \cite{zeghidour_soundstream_2021, defossez_high_2023, kumar_high_2023, wu_audiodec_2023, pia_nesc_2022} follow similar design patterns consisting of an end-to-end trained convolutional encoder-decoder architecture with quantization at the bottleneck. The work horse for \acrshort{sota} neural speech coding is the \acrfull{gan} training paradigm %\cite{goodfellow_generative_2014}
which enjoyed great popularity, especially in the computer vision field where it has been applied to various image generation tasks \cite{brock_large_2019, karras_progressive_2018, karras_style_2019}. Recently, \acrlong{dm}s (\acrshort{dm}s) surpassed \acrshort{gan} performance on image generation \cite{dhariwal_diffusion_2021}.
Moreover, \acrshort{dm}s do not suffer from the well-known training issues of \acrshort{gan}s such as mode collapse and vanishing gradients, making them an attractive alternative to \acrshort{gan}s for generative tasks.

In the audio domain, \acrshort{dm}s have been applied to several fields, including audio synthesis \cite{chen_wavegrad_2021, kong_diffwave_2021, huang_fastdiff_2022} and audio denoising \cite{serra_universal_2022, tian_diffusion-based_2023, welker_speech_2022}. Recently, the first diffusion-based audio and speech codecs started to emerge and showed promising results: LaDiffCodec (\LDC{}) \cite{yang_generative_2024} upsamples and dequantizes low-bitrate EnCodec (\EC{}) \cite{defossez_high_2023} tokens with a latent \acrshort{dm} to produce a continuous latent that is decoded by an \EC{} decoder pretrained on continuous input data, i.e., without quantization.
Similarly, Multi-Band Diffusion (\MBD{}) \cite{roman_discrete_2023} conditions on the \EC{} latent but directly generates decoded waveforms by independently processing different frequency bands. %The motivation for this approach lies in the fact that diffusion models tend to generate low-frequency content before high-frequency content, and are thus prone to accumulating errors in the high frequencies at inference time.

New speech and audio coding approaches have been proposed by combining iterative sampling methods, such as \acrshort{dm}s or \acrfull{cfm} models, with other advanced deep learning techniques, notably semantic embeddings: SemantiCodec \cite{liu_semanticodec_2024} and MuCodec \cite{xu_mucodec_2024} target the ultra-low bitrate regime ($0.3$ to $1.4$ kbps) by combining, sematic embeddings, \acrshort{dm} or \acrshort{cfm}, pretrained \acrfull{vae} for mel-spectrogram reconstruction and vocoding. %The architecture consists of two \acrshort{lm}-based encoders that separately capture semantic and acoustic features, a latent \acrshort{dm} that conditions on the quantized features of the LM encoders to generate a latent representation, and a VAE-based vocoder that synthesizes the audio signal from the latent. MuCodec \cite{xu_mucodec_2024} is based on four sequentially-trained models: An encoder modeling a discrete representation, a conditional flow matching latent model, a pretrained Mel-VAE a mel-based vocoder \cite{kong_hifi-gan_2020}.
In this paper, we exclude the computationally complex approaches that employ semantic embeddings, which seem to be crucial when specifically targeting ultra-low bitrates.

%\subsection{Motivation and Outline}

Diffusion-based \acrshort{nsc}s have not yet attracted the research interest that would be expected from the success of \acrshort{dm}s in image and audio synthesis and are underrepresented in the literature compared to \acrshort{gan}s. Furthermore, the design space of \acrshort{dm}-based speech codecs has not been explored systematically: the
existing diffusion-based codecs \cite{yang_generative_2024, xu_mucodec_2024, liu_semanticodec_2024, roman_discrete_2023} are the result of arguably one of the most substantial design choices, namely the conditioning and output domains of the \acrshort{dm}, whose impact on speech generation quality is unclear. %For example, both Multi-Band Diffusion and LaDiffCodec condition on the latent representation of an EnCodec model. One of the main differences between the two models is the output domain of the \acrshort{dm}: waveform for Multi-Band Diffusion and latent for LaDiffCodec.

We systematically explore the design space of diffusion-based codecs by three contributions:

\begin{enumerate}
\item We propose a categorization based on the \acrshort{dm} conditioning/output domain, where we consider waveform (\wav{}), mel-spectrogram (\mel{}) and latent space (\lat{}) representations.

\item All possible combinations of conditioning/output domain pairs from the representations mentioned above are systematically explored, except for using \wav{} as conditioning, which is infeasible for low bitrates due to the high dimensionality of time-domain signals.

\item We evaluate the proposed models and compare them to \acrshort{gan}-based and \acrshort{dm}-based baselines.
\end{enumerate}

%The main contribution of this paper is to evaluate the performance of several diffusion-based codecs by fully exploring the conditioning/output categorization outlined above. To that end, we
%\begin{enumerate}
 %   \item Train one configuration for each conditioning/output pair at 1.5 and 3 kbps. Select the best performing configuration and then train an additional model at 6 kbps.
    
  %  \item Compare against VQ-GAN (EnCodec) and diffusion-based baselines (Multi-Band Diffusion, LaDiffCodec) at 1.5, 3 kbps and 6 if available
    
   % \item Compare against retrained SQ-GAN baselines (\acrshort{sq} EnCodec, \acrshort{sq} HiFiGAN), 1.5, 3, 6 kbps.

%\end{enumerate}

\section{Diffusion Model-based Speech Codecs}
\label{section:diffusion}
\begin{figure}
    \centering
    \input{tikz/fig1}
    \caption{Sampling scheme of the proposed \acrshort{dm}-based \acrshort{nsc}s. The encoder can be fixed (\mel{}) or learned (\lat{}).}
    \label{fig:dm_block_diagram}
\end{figure}
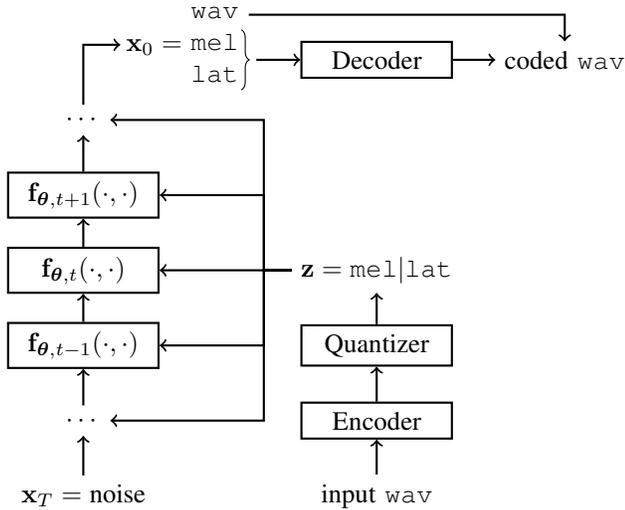

In the following, we give a brief overview that covers the main principles of generative \acrshort{dm}s \cite{sohl-dickstein_deep_2015} for neural speech coding. Notably, there exist two formulations of \acrshort{dm}s: a continuous-time description based on \acrlong{sde}s (\acrshort{sde}s) and a discrete-time framework based on Markov Chains which is typically referred to as \acrlong{ddpm}s (\acrshort{ddpm}s). We refer the reader to \cite{luo_understanding_2022} and \cite{song_score_2021 %,song_maximum_2021
} for further details regarding \acrshort{ddpm}s and \acrshort{sde}-based \acrshort{dm}s, respectively.

\acrshort{dm}s for neural speech coding model a stochastic process transforming speech samples (or derived representations such as mel-spectra or latent embeddings) $\mathbf{x}_0 \sim p_{\text{data}}(\mathbf{x})$ into standard Gaussian noise samples $\mathbf{x}_T \sim \mathcal{N}(\mathbf{0},\mathbf{I})$. This transformation, typically called the forward diffusion process, may be expressed by
\begin{equation}
    \mathbf{x}_t = a_t \mathbf{x}_0 + b_t \bm{\epsilon},\ \text{with}\ \bm{\epsilon}\sim\mathcal{N}(\mathbf{0},\mathbf{I}),
\label{eq:forwardDiff}
\end{equation}
where $t$ is a time index (not related to the speech signal but to the diffusion process), which can be discrete $t\in \{1,\dots, T\}$ (\acrshort{ddpm}s) or continuous $t\in[0,T]$ (\acrshort{sde}-based \acrshort{dm}s).
Here, $a_t, b_t \in \mathbb{R}_{\geq 0}$ are time-dependent coefficients chosen according to a user-defined noise schedule such that $\mathbf{x}_T\sim\mathcal{N}(\mathbf{0},\mathbf{I})$.  

\acrshort{dm}s are trained to reverse the forward diffusion process by transforming samples from the standard Gaussian noise prior $\mathbf{x}_T\sim\mathcal{N}(\mathbf{0},\mathbf{I})$ into speech samples, i.e., samples following the data distribution $\mathbf{x}_0 \sim p_{\text{data}}(\mathbf{x})$. This so-called reverse diffusion process is modeled by a \acrfull{dnn}-based function $\mathbf{f}_{\bm{\theta}, t}$ that is parameterized by $\bm{\theta}$ and is dependent on the time step $t$. The \acrshort{dnn} $\mathbf{f}_{\bm{\theta}, t}$ is trained to estimate the speech sample $\mathbf{x}_0$ from a noisy version of it $\mathbf{x}_t$. Equivalently \cite{luo_understanding_2022}, the \acrshort{dnn} can be trained to either predict the noise added in the forward process, i.e., $\bm{\epsilon}$ in Eq.~\eqref{eq:forwardDiff}), or to estimate the `score' function, i.e., $\nabla_{\mathbf{x}_t} \log{p(\mathbf{x}_t)}$. For generating new speech samples, we sample from the standard Gaussian noise prior $\mathbf{x}_T\sim\mathcal{N}(\mathbf{0},\mathbf{I})$ and apply the following steps iteratively until $t=0$
\begin{align}
    \mathbf{x}_{t} &\leftarrow \mathbf{f}_{\bm{\theta}, t}(\mathbf{x}_{t}) + c_t \bm{\epsilon},\ \text{with}\ \bm{\epsilon}\sim\mathcal{N}(\mathbf{0},\mathbf{I})\\
    t & \leftarrow t - \Delta t.
\end{align}
Here, $c_t\in \mathbb{R}_{\geq 0}$ denotes a coefficient that increases in $t$ depending on the noise schedule and $\Delta t$ denotes the discretization step for the diffusion time axis ($\Delta t = 1$ in the discrete-time case). Hence, the noisy speech sample $\mathbf{x}_{t}$ is denoised step by step in the reverse diffusion process.

\acrshort{nsc}s aim to generate speech signals that are natural, i.e., signals that follow the data distribution $\mathbf{x}_0 \sim p_{\text{data}}(\mathbf{x})$, while also sounding as similar as possible to the signal to be transmitted.
To that end, we implement a control mechanism to guide the sampling procedure, by providing the \acrshort{dm} with conditioning information $\conditioning$ both at training and inference time, i.e., $\mathbf{f}_{\bm{\theta}, t}(\mathbf{x}_{t}, \conditioning)$. As commonly done in \acrshort{nsc}s, we employ quantization methods to learn compact discrete representations that are suitable for transmission, which will also be the basis for the mentioned conditioning information $\conditioning$. Several \acrshort{nsc}s \cite{defossez_high_2023, kumar_high_2023, zeghidour_soundstream_2021, pia_nesc_2022, wu_audiodec_2023} use \acrfull{rvq} \cite{chen_approximate_2010}, i.e., a cascade of vector quantizers each encoding the residual of its predecessor. However, \acrshort{rvq} suffers from well-known drawbacks: it requires reinitialization and decision procedures to avoid underutilized codevectors (codebook collapse), careful hyperparameter tuning, and extra training losses. \acrfull{sq} addresses the shortcomings of \acrshort{rvq} \cite{mentzer_finite_2023} and has been successfully applied in the image domain \cite{balle_nonlinear_2021}. For training, \acrshort{sq} can be approximated by noise addition (`NoiseSQ') which simplifies training while achieving results comparable to \acrshort{rvq}, as shown in \cite{brendel_simple_2024}. Due to the noise addition, training NoiseSQ end-to-end with a neural codec yields a smoother latent distribution, which is desirable for latent space modeling with generative models.

\section{Experimental Setup}

\subsection{Model Design}

We investigate the choice of the \acrshort{dm} conditioning and output domains for diffusion-based \acrshort{nsc}s. Waveform (\wav{}), mel-spectrogram (\mel{}) and latent embeddings (\lat{}) are popular representations of speech signals and are considered as output domain choices, whereas \mel{} and \lat{} are examined for \acrshort{dm} conditioning. Since \acrshort{nsc}s require discrete conditioning information, we always assume that the mel/latent representation used for conditioning is quantized, e.g., with \acrshort{sq}. % Conditioning on quantized waveforms is unfeasible due to the very high bitrate requirement needed to transmit the conditioning information.
Based on this conceptual framework, we identify six conditioning/output configurations: \mTw{}, \lTw{}, \mTm{}, \lTm{}, \mTl{}, \lTl{}. To the best of our knowledge, only two out of the six model designs have been already explored in the literature: \MBD{} \cite{roman_discrete_2023} and \LDC{} \cite{yang_generative_2024}. \MBD{} belongs to the \lTw{} category since the \acrshort{dm} conditions on the \EC{} latent and outputs waveforms, whereas \LDC{} is a \lTl{} model. In general, we refer to \mTw{} and \lTw{} as `waveform diffusion' approaches, since the output of the \acrshort{dm} is a signal in the time domain. Similarly, \mTm{} and \lTm{} are grouped under `mel diffusion' and \mTl{} and \lTl{} under `latent diffusion'. Fig.~\ref{fig:dm_block_diagram} provides a schematic overview of speech generation with the proposed design setup using the terminology introduced in Sec.~\ref{section:diffusion}. Note that the quantizer can include learnable projections, e.g., as in \acrshort{sq} \cite{brendel_simple_2024}.

The choice of the \acrshort{dm} output domain has several implications, e.g., waveform diffusion is more computationally complex compared to mel or latent diffusion. The latter approaches, on the other hand, require an additional model, a mel vocoder or a latent decoder for mel-spectra and latent, respectively. Mel diffusion provides better interpretability compared to latent diffusion and does not require a neural encoder. On the other hand, we expect a latent representation specifically learned for coding to be more powerful and thus achieve better results compared to the very generic speech representation by mel-spectra.

 In this paper, the focus is primarily on investigating which choice of conditioning and output domain yields the best performance for neural speech coding. To that end, we employ well-known \acrshort{dm} backbones from audio synthesis. In order to make the comparison as fair as possible, we use the same \acrshort{dnn} architecture for latent and mel diffusion, which is possible due to the similar dimensionality of these speech data representations, and use a different model only for waveform diffusion.
\DW{} \cite{kong_diffwave_2021}, a \acrshort{sota} diffusion-based vocoder, is chosen as the main building block to realize \mTw{} and \lTw{}. Similarly, we use GradTTS \cite{popov_grad-tts_2021}, which, in addition to text-to-speech, has also been applied for speech denoising \cite{tian_diffusion-based_2023}, for \mTm{}, \mTl{}, \lTm{} and \lTl{}. Following \cite{roman_discrete_2023, yang_generative_2024}, we leverage \EC{} as GAN-based baseline and to learn an expressive latent representation to be used as conditioning for the \acrshort{dm}s. We consider BigVGAN-base \cite{lee_bigvgan_2022} and HiFiGAN V1 \cite{kong_hifi-gan_2020} as vocoder models. Table~\ref{tab:models} gives an overview of these models as they were proposed in the literature.

\subsection{Training and Evaluation}
\label{subsection:trainingAndEvaluation}

A general training setup applies to all the models: the models were trained for 1 million steps on clean speech signals with a fixed segment length of 1 second (convergence of the models has been confirmed for each training). The training data comprise the LibriTTS \cite{zen_libritts_2019} and VCTK \cite{yamagishi_vctk_2019} datasets at 16 kHz. %, for a total of approximately 300 hours.
For each model architecture, we followed the recommended training hyperparameter choices (optimizer parameters, batch size, noise schedule, etc.) indicated in the respective publications. %After training, the models were confirmed to have converged by monitoring the loss values.

An internal English test set consisting of 28 speech signals, 14 female and 14 male utterances of 8 seconds duration, was used for assessing the models' performance based on objective and subjective evaluation. 
For objective evaluation, we used \visqol{} \cite{hines_visqol_2012} and \scoreq{} \cite{ragano_scoreq_2024}, a perceptual-based and a learning-based speech quality objective metric, respectively. Details about subjective \acrfull{lt} evaluation are given in Sec.~\ref{subsection:listeningTests}.

\subsection{Preliminary Experiments}
\label{subsection:preliminaryExperiments}

\paragraph{Mel-spectrogram diffusion} Vocoders %operating on mel inputs
are essential for the \acrshort{dm} configurations that generate mel-spectra, namely \mTm{} and \lTm{}. Thus, we run a set of preliminary experiments to choose the vocoder architecture and hop size.
We retrained BigVGAN-base \cite{lee_bigvgan_2022} and HiFiGAN V1 \cite{kong_hifi-gan_2020} in various configurations following the official implementation\footnote{\url{https://github.com/NVIDIA/BigVGAN}} and found that they showed similar performance. %HiFiGAN V1 showed similar performance than BigVGAN, despite its lower complextiy.
Furthermore, larger hop sizes seem to yield more robust results when the vocoders take degraded mel-spectra as input (which represents a training test mismatch). Thus, we selected HiFiGAN V1 with a hop size of 256 as the vocoder for the following experiments.

%with hop size of 128 and 256. In a set of preliminary experiments, the two models did not show significant differences in terms of performance. Therefore, we selected HiFiGAN as our vocoder model due to its lower complexity compared to BigVGAN. Furthermore, we observed that the vocoders trained with a hop size of 256 were more robust when presented to degraded mel-spectra (outputs of pretrained \acrshort{sq} models with reconstruction loss). Thus, we selected 256 as the hop size of the vocoder.

\paragraph{Latent diffusion} In order to choose the quantizer model, we trained from scratch six \EC{} models with either \acrshort{sq} or \acrshort{rvq}, at 1.5, 3 and 6 kbps. In this and all subsequent experiments, \acrshort{sq} was trained as an autoencoder with noise addition at the bottleneck, as for NoiseSQ in \cite{brendel_simple_2024}. We followed the official \EC{} model implementation\footnote{\url{https://github.com/facebookresearch/encodec}} and training hyperparameters \cite{defossez_high_2023}, except for modifying the hop size and latent dimension of the original model from 320 and 128 to 256 and 80 respectively, to match the hop size of and number of mel-bands of HiFiGAN, thereby allowing for easier comparison. The modified \EC{} model has downsampling/upsampling ratios of $\left[8,4,4,2\right]$ and $\left[2,4,4,8\right]$ respectively. As \acrshort{sq} performed at least as good as \acrshort{rvq} for all bitrates, we chose \acrshort{sq} as the quantizer for the following experiments. This also allows for learning smooth latent representations for latent \acrshort{dm}s as argued above.

%For all bitrates, \acrshort{sq} performed on par or better than RVQ. Moreover, in contrast to \acrshort{rvq}, \acrshort{sq} is known to produce a smooth latent representation, which is better suited to be learned by diffusion models. For these reasons, we chose to use \acrshort{sq} in all our experiments.

\begin{table}[h]
    \centering
    \resizebox{0.45\textwidth}{!}{%
    \begin{tabular}{lcccccc}
        \toprule
        \textbf{Model} & \textbf{Framework} & \textbf{Input/Cond.} & \textbf{Out} & \textbf{Param. (M)} & \textbf{GMACs} \\
        \midrule
        EnCodec  & GAN & wav & wav & 14.42 &  1.66 \\
        HiFiGAN  & GAN & mel & wav & 13.93 & 19.35 \\
        BigVGAN  & GAN & mel & wav & 13.94 & 19.72 \\
        DiffWave & DM  & mel & wav &  2.66 & 41.78 \\
        GradTTS  & DM  & mel & mel & 91.41 & 16.57 \\

        \bottomrule
    \end{tabular}
    }
    \caption{Overview of models from the literature. The complexity values (GMACs) refer to a single step forward-pass.}
    \label{tab:models}
\end{table}
\vspace{-10pt}

% \begin{table}[h]
%     \centering
%     \begin{tabular}{lcc}
%         \toprule
%         \textbf{Design} & \textbf{Models} \\
%         \midrule
%         \mTw{} & Mel $\rightarrow$ \acrshort{sq} $\rightarrow$ DiffWave  \\
%         \lTw{} & EnCodec Encoder $\rightarrow$ \acrshort{sq} $\rightarrow$ DiffWave  \\

%         \mTm{} & Mel $\rightarrow$ \acrshort{sq} $\rightarrow$ GradTTS $\rightarrow$ HiFiGAN  \\
%         \lTm{} & EnCodec Encoder $\rightarrow$ \acrshort{sq} $\rightarrow$ GradTTS $\rightarrow$ HiFiGAN  \\

%         \mTl{} & Mel $\rightarrow$ \acrshort{sq} $\rightarrow$ GradTTS $\rightarrow$ EnCodec Decoder  \\
%         \lTl{} & EnCodec Encoder $\rightarrow$ \acrshort{sq} $\rightarrow$ GradTTS $\rightarrow$ EnCodec Decoder \\

%         \bottomrule
%     \end{tabular}
%     \caption{Overview of proposed designs.}
%     \label{tab:designs}
% \end{table}

\begin{table}[h]
    \centering
    \begin{tabular}{lcccc}
        \toprule
        \textbf{Design} & \textbf{Encoder} & \textbf{DM} & \textbf{Decoder/Vocoder} \\
        \midrule
        
        \mTw{} & Mel & DiffWave & \tablePlaceholder{} \\
        \lTw{} & EnCodec & DiffWave & \tablePlaceholder{}  \\

        \mTm{} & Mel & GradTTS & HiFiGAN  \\
        \lTm{} & EnCodec  & GradTTS & HiFiGAN  \\

        \mTl{} & Mel & GradTTS & EnCodec  \\
        \lTl{} & EnCodec & GradTTS & EnCodec  \\

        %QGAN & Mel & \tablePlaceholder{}  & HiFiGAN  \\
        %EnCodec & Encodec & \tablePlaceholder{} & Encodec \\

        \bottomrule
    \end{tabular}
    \caption{Overview of proposed designs.}
    \label{tab:designs}
\end{table}

\subsection{Experiments}
\label{subsection:experiments}

\begin{figure*}[t]
\begin{minipage}{0.67\textwidth}
    \centering
    \begin{subfigure}[t]{0.33\linewidth}
        \centering
        \resizebox{\linewidth}{\highfigheight{}}{\input{tikz/exp1_scoreq}}
    \end{subfigure}%
    \begin{subfigure}[t]{0.33\linewidth}
        \centering
        \resizebox{\linewidth}{\highfigheight{}}{\input{tikz/exp2_scoreq}}
    \end{subfigure}%
    \begin{subfigure}[t]{0.33\linewidth}
        \centering
        \resizebox{\linewidth}{\highfigheight{}}{\input{tikz/exp3_scoreq}}
    \end{subfigure}%
    %\hfill
    \vspace{0mm}
    \begin{subfigure}[t]{0.33\linewidth}
        \centering
        \resizebox{\linewidth}{\firstlowfigheight{}}{\input{tikz/exp1_visqol}}
        \caption{}
        \label{fig:exp1}
    \end{subfigure}%
    \begin{subfigure}[t]{0.33\linewidth}
        \centering
        \resizebox{\linewidth}{\lowfigheight{}}{\input{tikz/exp2_visqol}}
        \caption{}
        \label{fig:exp2}
    \end{subfigure}%
    \begin{subfigure}[t]{0.33\linewidth}
        \centering
        \resizebox{\linewidth}{\lowfigheight{}}{\input{tikz/exp3_visqol}}
        \caption{}
        \label{fig:exp3}
    \end{subfigure}
\caption{Objective evaluation of Exp. 1 (a), 2 (b) and 3 (c). (b) shows retrained models on the left of the dashed line, pretrained ones on the right, with \acrshort{gan} and \acrshort{dm} models colored in red and blue respectively.}
\end{minipage}
\hfill
\begin{minipage}{0.32\textwidth}
    \centering
    \resizebox{\linewidth}{\figHeight{}}
    {\input{tikz/lt1}}
    \caption{P.808 DCR test results (including 15 listeners) comparing the proposed \acrshort{dm}-based \acrshort{nsc}s at 3 kbps.}
    \label{fig:LT1}
\end{minipage}
\end{figure*}

\paragraph{Exp. 1 - Evaluation of proposed designs}
\label{paragraph:exp1}

We assess the impact of the \acrshort{dm} conditioning/output design choice by evaluating the proposed diffusion-based \acrshort{nsc} configurations. In this experiment, all \acrshort{dm} models condition on a discrete representation quantized at 3 kbps. For \mTw{}, \mTm{} and \mTl{}, the \acrshort{dm} and \acrshort{sq} are trained end-to-end. We found it beneficial to use an additional reconstruction loss (sum of $L_{1}$ and $L_{2}$ losses) to train \acrshort{sq}. \lTw{}, \lTm{} and \lTl{} are trained using the quantized latent embeddings of a pretrained \EC{} model as conditioning. The jointly pretrained \EC{} encoder and quantizer are frozen when training the \acrshort{dm}s. For the latent diffusion configurations, the ceiling quality is determined by the decoder. Thus, we pretrain a high-bitrate \EC{} with \acrshort{sq} at 8 kbps, which achieves very good speech quality, and subsequently train \mTl{} and \lTl{} to generate the latent embeddings of the high-bitrate \EC{}. The main difference of our approach compared to \LDC{} \cite{yang_generative_2024} is that the \EC{} model, whose latent space is to be learned by the \acrshort{dm}, is trained with \acrshort{sq} to enforce a smooth latent representation thereby facilitating the \acrshort{dm} generative task.%We believe that our approach is more promising, due to the regularization effect of \acrshort{sq}, which enforces a smoother latent representation that should be easier to model with a \acrshort{dm}. 

Table~\ref{tab:designs} provides an overview of the proposed designs.

\paragraph{Exp. 2 - Best proposed design vs baselines}

Here, we compare the best performing configuration from Exp. 1 to \acrshort{sota} \acrshort{dm} and \acrshort{gan}-based baselines at 3 kbps. 
The following baselines are included:
\begin{itemize}
    \item Pretrained \EC{} (\acrshort{gan}-based) and \MBD{} (\acrshort{dm}-based) (checkpoints and inference code available\footnote{\url{https://huggingface.co/facebook/multiband-diffusion}}). Both models were trained for coding general audio at 24 kHz with variable bitrate (1.5, 3, 6 kbps for \MBD{}, 1.5, 3, 6 12, 24 kbps for \EC{}). Since our models are trained for a single fixed bitrate and only on speech, we expect better results than these baselines.
    \item Pretrained \LDC{}, 3 kbps model trained on Librispeech \cite{panayotov_librispeech_2015} (clean-100) at 16 kHz (checkpoint and inference code available\footnote{\url{https://github.com/haiciyang/LaDiffCodec}}).
    \item Retrained \EC{} with \acrshort{sq} at 3 kbps (see Sec.~\ref{subsection:preliminaryExperiments}).
    \item Retrained HiFiGAN V1 with \acrshort{sq} at 3 kbps (quantizer and vocoder trained end-to-end). Similarly to \mTw{}, \mTm{} and \mTl{}, \acrshort{sq} is trained with a reconstruction loss. We refer to this baseline as \QGAN{}.
\end{itemize}

It is worth to emphasize that mel and latent diffusion are evaluated without fine-tuning the decoder/vocoder model, i.e., with a `non-matched' condition. Intuitively, fine-tuning is expected to improve performance as the decoder/vocoder can learn to adapt to the input generated by the \acrshort{dm}. Since \mTm{} will prove to be the best performer of Exp. 1, we include in the comparison the `matched' condition by fine-tuning the pretrained HiFiGAN vocoder on VCTK \cite{yamagishi_vctk_2019}. % 70000 steps.

\paragraph{Exp. 3 - Best performers at different bitrates}

\mTm{}, the best performer of Exp. 1, is compared to the best performing baselines of Exp. 2, namely the retrained \EC{} and \QGAN{}. All models are evaluated at 1.5, 3, and 6 kbps.

\subsection{Listening Tests}
\label{subsection:listeningTests}

To support and validate the objective metrics evaluation, we run two subjective \acrshort{lt}s with Degradation Category Ratings (DCR) following the ITU-T P.808 principles \cite{ITU-T_P808} on the test set described in Sec.~\ref{subsection:trainingAndEvaluation}. The first \acrshort{lt} includes the results of Exp. 1, while the second \acrshort{lt} comprises all results of Exp. 2 and 3.

\section{Experimental Results}

In Exp. 1, the proposed diffusion-based \acrshort{nsc} designs are compared at a bitrate of 3 kbps. Based on the objective metrics, the best performing configuration is \mTm{} as shown in Fig.~\ref{fig:exp1}. In general, \scoreq{} evaluates mel diffusion as the best paradigm, followed by waveform diffusion and latent diffusion, a conclusion which is supported by the results of the first \acrshort{lt} shown in Fig.~\ref{fig:LT1}, whereas \visqol{} scores show a less definite trend.

Fig.~\ref{fig:exp2} presents the results of Exp. 2, showing that \mTm{} significantly outperforms the pretrained baselines, while being slightly worse than the retrained \QGAN{} and \EC{} models. However, we observe that, as expected, fine-tuning improves \mTm{} performance, and that the fine-tuned \mTm{} achieves better or on-par results to the retrained \acrshort{gan} models.

Fig.~\ref{fig:exp3} depicts the outcome of Exp. 3, where the best performers of Exp. 2, \mTm{}, the retrained \EC{} and \QGAN{}, are evaluated at 1.5, 3 and 6 kbps. Overall, we find that \mTm{} without fine-tuning (`non-matched' condition) achieves a comparable performance to \QGAN{} (`matched' condition), but performs worse than \EC{}. However, fine-tuning \mTm{} significantly reduces the performance gap to \EC{}, even yielding better scores for the 6 kbps models.

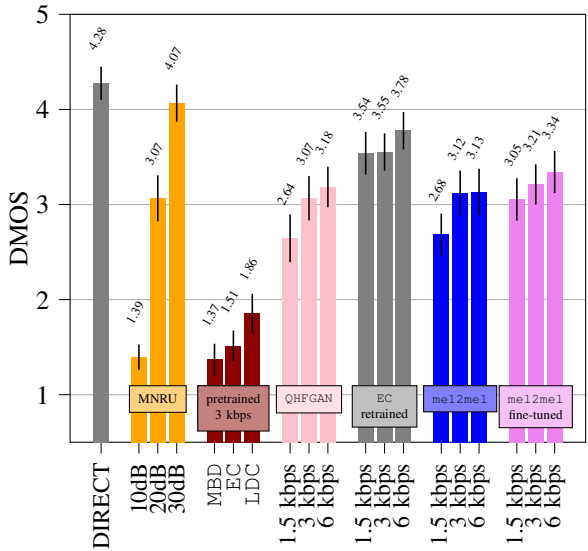
\begin{figure}
    \vspace{-5mm}\centering
    %\resizebox{\linewidth}{8cm}
    {\input{tikz/lt2}}
    \caption{P.808 DCR test results (including 19 listeners) comparing \mTm{} to \acrshort{gan} and \acrshort{dm} baselines.}
    \label{fig:LT2}
\end{figure}

Fig.~\ref{fig:LT2} shows the outcome of a second \acrshort{lt} that comprises the models from Exp. 2 and 3. Consistent with the objective metrics evaluation, we observe that the retrained models outperform the pretrained ones by a large margin. Moreover, \mTm{}, both with and without fine-tuning is shown to achieve better or comparable ratings compared to \QGAN{}. The retrained \EC{} appears to be the best performing model, which is in line with the objective evaluation.

\section{Conclusion}

In this paper, we explored the design space of diffusion-based \acrshort{nsc}s by investigating which conditioning/output configuration produces the best speech quality. The proposed designs were compared to \acrshort{sota} \acrshort{gan} and \acrshort{dm} baselines through objective and subjective evaluation. According to our findings, the best proposed design was \mTm{}, where a \acrshort{dm} generates enhanced mel-spectra from quantized mel-spectra. \mTm{} performed better than other \acrshort{dm}-based baselines proposed in the literature, but fails to improve on the results of \EC{}, a \acrshort{sota} \acrshort{gan}-based codec.

%\section*{Acknowledgment}

%\renewcommand{\IEEEbibitemsep}{6pt}
\bibliographystyle{IEEEtran}
\bibliography{IEEEabrv, bib/clean_bib_no_pages
}
%,bib/bib_ae, bib/bib_codecs, bib/bib_datasets, bib/bib_ddpm, bib/bib_gan, bib/bib_github, bib/bib_metrics, bib/bib_misc, bib/bib_quant, bib/bib_repr, bib/bib_software}

\end{document}

%% file: 00_Commands.tex
\newcommand{\readableBegin}
{\begin{mdframed}[linewidth=1.5,linecolor=red, innertopmargin=-.1em, innerbottommargin=0em, topline=false, rightline=false,bottomline=false]}
\newcommand{\readableEnd}{\end{mdframed}}

\newcommand{\boldFaceCommand}[1]{\mathbf{#1}}

\newcommand{\encodecModel}{\mathtt{EC}}

\newcommand{\brac}[1]{\left( #1 \right)}

\newcommand{\ladiffModel}{\mathtt{LD}}

\newcommand{\pretrained}{\mathtt{PT}}

\newcommand{\DW}{DiffWave}

\newcommand{\visqol}{ViSQOL}

\newcommand{\scoreq}{SCOREQ}

\newcommand{\innermid}{\nonscript\;\delimsize\vert\nonscript\;}
\newcommand{\activatebar}{%
  \begingroup\lccode`\~=`\|
  \lowercase{\endgroup\let~}\innermid 
  \mathcode`|=\string"8000
}

\newcommand{\conditioning}{\boldFaceCommand{z}}

\newcommand{\wav}{\texttt{wav}}
\newcommand{\mel}{\texttt{mel}}
\newcommand{\lat}{\texttt{lat}}

\newcommand{\XtoY}{\texttt{2}}

\newcommand{\mTw}{\mel{}\XtoY{}\wav{}}
\newcommand{\lTw}{\lat{}\XtoY{}\wav{}}

\newcommand{\mTm}{\mel{}\XtoY{}\mel{}}
\newcommand{\lTm}{\lat{}\XtoY{}\mel{}}

\newcommand{\mTl}{\mel{}\XtoY{}\lat{}}
\newcommand{\lTl}{\lat{}\XtoY{}\lat{}}

\newcommand{\tablePlaceholder}{--}

\newcommand{\EC}{\texttt{EC}}

\newcommand{\MBD}{\texttt{MBD}}
\newcommand{\LDC}{\texttt{LDC}}
\newcommand{\QGAN}{\texttt{QHFGAN}}

\newcommand{\highfigheight}{2.6cm}
\newcommand{\lowfigheight}{3.6cm}
\newcommand{\firstlowfigheight}{3.65cm}

\newcommand{\figHeight}{6.4cm}

\newcommand{\plotTickFont}{\Large}

%% file: 00_Glossary.tex
\makeglossaries

\newacronym{gan}{GAN}{Generative Adversarial Network}

\newacronym{gm}{GM}{Generative Models}

\newacronym{dm}{DM}{Diffusion Model}

\newacronym{ae}{AE}{Autoencoder}

\newacronym{vae}{VAE}{Variational Autoencoder}

\newacronym{ld}{LD}{Latent Diffusion}

\newacronym{ec}{EC}{EnCodec}

\newacronym{wg}{WG}{WaveGrad}

\newacronym{dw}{DW}{DiffWave}

\newacronym{hf}{HF}{HiFiGAN}

\newacronym{dmse}{DMSE}{Diffusion Mel-spectrogram Enhancement}

\newacronym{mbd}{MBD}{Multi-band Diffusion}

\newacronym{tdd}{TDD}{Time-domain Diffusion}

\newacronym{msd}{MSD}{Mel-spectrogram Diffusion}

\newacronym{eld}{ELD}{Encodec Latent Diffusion}

\newacronym{mae}{MAE}{Mean Absolute Error}

\newacronym{mse}{MSE}{Mean Squared Error}

\newacronym{msc}{MSC}{Mel-spectrogram Conditioner}

\newacronym{elc}{ELC}{Encodec Latent Conditioner}

\newacronym{elbo}{ELBO}{Evidence Lower Bound}

\newacronym{kld}{KLD}{Kullback-Leibler Divergence}

\newacronym{gelu}{GELU}{Gaussian Error Linear Unit}

\newacronym{sq}{SQ}{Scalar Quantization}

\newacronym{rvq}{RVQ}{Residual Vector Quantization}

\newacronym{ldc}{Latent Diffusion Codec}{LaDiffCodec}

\newacronym{lm}{LM}{Language Model}

\newacronym{dft}{DFT}{Discrete Fourier Transform}

\newacronym{stft}{STFT}{Short-Time Fourier Transform}

\newacronym{mac}{MAC}{Multiply-Accumulate}

\newacronym{stoi}{STOI}{Short-time Objective Intelligibility}

\newacronym{visqol}{ViSQOL}{Virtual Speech Quality Objective Listener}

\newacronym{nomad}{NOMAD}{Non-matching Audio Distance}

\newacronym{pesq}{PESQ}{Perceptual Evaluation of Speech Quality}

\newacronym{sota}{SOTA}{state-of-the-art}

\newacronym{nsc}{NSC}{Neural Speech Codec}

\newacronym{cfm}{CFM}{Conditional Flow Matching}

\newacronym{lt}{LT}{Listening Test}

\newacronym{dnn}{DNN}{Deep Neural Network}

\newacronym{sde}{SDE}{Stochastic Differential Equation}

\newacronym{ddpm}{DDPM}{Denoising Diffusion Probabilistic Model}

%% file: tikz/fig1.tex
\newcommand{\diffChainX}{0}
\newcommand{\condX}{3.9}
\newcommand{\baseY}{0}
\newcommand{\diffChainYShift}{1}
\usetikzlibrary{calc}

\tikzstyle{rect} = [rectangle, minimum width=1cm, minimum height=1cm, text centered, draw=none, fill=none]

\begin{tikzpicture}

    \node (noise) at (\diffChainX,\baseY){$\mathbf{x}_T=\text{noise}$};
    
    \node (dotsminus) at ($(noise)+(0, \diffChainYShift)$){$\cdots$};
    
    \node[draw, thick, minimum width=2cm, text centered] (tminus) at ($(dotsminus)+(0, \diffChainYShift)$){$\mathbf{f}_{\bm{\theta}, t-1}(\cdot,\cdot)$};
    
    \node[draw, thick, minimum width=2cm, text centered] (t) at ($(tminus)+(0, \diffChainYShift)$){$\mathbf{f}_{\bm{\theta}, t}(\cdot,\cdot)$};
    
    \node[draw, thick, minimum width=2cm, text centered] (tplus) at ($(t)+(0, \diffChainYShift)$){$\mathbf{f}_{\bm{\theta}, t+1}(\cdot,\cdot)$};
    
    \node (dotsplus) at ($(tplus)+(0, \diffChainYShift)$){$\cdots$};
    
    \node[draw=none, inner sep=2pt] (out) at ($(dotsplus)+(1.5, \diffChainYShift)$){$\mathbf{x}_0=\,\,$\parbox{1cm}{\wav{}\\\mel{}\\\lat{}}};

    \draw[decorate, decoration={brace, amplitude=3pt}, rotate=0] ($(out)+(0.6, 0.2)$) -- ($(out)+(0.6, -0.6)$) {};

    \node[draw, thick, minimum width=2cm, text centered] (dec) at ($(out)+(2.4, -0.2)$){Decoder};

    \node (wavPred) at ($(dec)+(2.5, 0)$){coded \wav{}};

    \draw[->, thick]($(out)+(0.7, 0.4)$) -| (wavPred);
    \draw[->, thick]($(out)+(0.8, -0.2)$) -- (dec);

    \draw[->, thick] (noise)--(dotsminus);
    \draw[->, thick](dotsminus)--(tminus);
    \draw[->, thick](tminus)--(t);
    \draw[->, thick](t)--(tplus);
    \draw[->, thick](tplus)--(dotsplus);
    \draw[->, thick](dotsplus)|-(out.west);

    \node (wav) at (\condX, \baseY){input \wav{}};
    
    \node[draw, thick, minimum width=2cm, text centered] (trafo) at (\condX, 1){Encoder};
    \node[draw, thick, minimum width=2cm, text centered] (quant) at (\condX, 2){Quantizer};

    \node (cond) at (\condX, 3){$\conditioning=\mel{}|\lat{}$};

    \coordinate (condKP) at ($(cond)+(-1.5,0)$){};

    \draw[->, thick](wav) -- (trafo);
    \draw[->, thick](trafo) -- (quant);
    \draw[->, thick](quant) -- (cond);

    \draw[->, thick](cond) -- (condKP) |-(dotsminus);
    \draw[->, thick](cond) -- (condKP) |-(tminus);
    \draw[->, thick](cond) -- (condKP) |-(t);
    \draw[->, thick](cond) -- (condKP) |-(tplus);
    \draw[->, thick](cond) -- (condKP) |-(dotsplus);

    \draw[->, thick](dec) -- (wavPred);

\end{tikzpicture}

%% file: tikz/exp1_scoreq.tex
% This file was created with tikzplotlib v0.10.1.
\begin{tikzpicture}

\definecolor{darkgray176}{RGB}{176,176,176}
\definecolor{forestgreen4416044}{RGB}{44,160,44}
\definecolor{steelblue31119180}{RGB}{31,119,180}

\begin{axis}[
tick align=outside,
tick pos=left,
x grid style={darkgray176},
xmajorgrids,
xmin=-0.5, xmax=5.5,
xlabel={},
xticklabels={},
y grid style={darkgray176},
ylabel={$\longleftarrow$ ~ \scoreq{}},
ylabel style={yshift=-10pt},
ymajorgrids,
ymin=0.10,
ymax=1.3,
ytick style={color=black}
]
\addplot [ultra thick, steelblue31119180]
table {%
-0.4 0.649819523096085
0.4 0.649819523096085
0.4 0.841080546379089
-0.4 0.841080546379089
-0.4 0.649819523096085
};
\addplot [ultra thick, steelblue31119180]
table {%
0 0.649819523096085
0 0.555688440799713
};
\addplot [ultra thick, steelblue31119180]
table {%
0 0.841080546379089
0 0.978109419345856
};
\addplot [ultra thick, steelblue31119180]
table {%
-0.2 0.555688440799713
0.2 0.555688440799713
};
\addplot [ultra thick, steelblue31119180]
table {%
-0.2 0.978109419345856
0.2 0.978109419345856
};
\addplot [ultra thick, steelblue31119180]
table {%
0.6 0.491337209939957
1.4 0.491337209939957
1.4 0.619917154312134
0.6 0.619917154312134
0.6 0.491337209939957
};
\addplot [ultra thick, steelblue31119180]
table {%
1 0.491337209939957
1 0.388637751340866
};
\addplot [ultra thick, steelblue31119180]
table {%
1 0.619917154312134
1 0.780057311058044
};
\addplot [ultra thick, steelblue31119180]
table {%
0.8 0.388637751340866
1.2 0.388637751340866
};
\addplot [ultra thick, steelblue31119180]
table {%
0.8 0.780057311058044
1.2 0.780057311058044
};
\addplot [ultra thick, steelblue31119180]
table {%
1.6 0.387603476643562
2.4 0.387603476643562
2.4 0.51121823489666
1.6 0.51121823489666
1.6 0.387603476643562
};
\addplot [ultra thick, steelblue31119180]
table {%
2 0.387603476643562
2 0.301101207733154
};
\addplot [ultra thick, steelblue31119180]
table {%
2 0.51121823489666
2 0.692492306232452
};
\addplot [ultra thick, steelblue31119180]
table {%
1.8 0.301101207733154
2.2 0.301101207733154
};
\addplot [ultra thick, steelblue31119180]
table {%
1.8 0.692492306232452
2.2 0.692492306232452
};
\addplot [black, mark=o, mark size=3, mark options={solid,fill opacity=0,draw=steelblue31119180}, only marks]
table {%
2 0.726335167884827
};
\addplot [ultra thick, steelblue31119180]
table {%
2.6 0.45147218555212
3.4 0.45147218555212
3.4 0.600991755723953
2.6 0.600991755723953
2.6 0.45147218555212
};
\addplot [ultra thick, steelblue31119180]
table {%
3 0.45147218555212
3 0.331970274448395
};
\addplot [ultra thick, steelblue31119180]
table {%
3 0.600991755723953
3 0.761409401893616
};
\addplot [ultra thick, steelblue31119180]
table {%
2.8 0.331970274448395
3.2 0.331970274448395
};
\addplot [ultra thick, steelblue31119180]
table {%
2.8 0.761409401893616
3.2 0.761409401893616
};
\addplot [black, mark=o, mark size=3, mark options={solid,fill opacity=0,draw=steelblue31119180}, only marks]
table {%
3 0.869396209716797
};
\addplot [ultra thick, steelblue31119180]
table {%
3.6 1.02589786052704
4.4 1.02589786052704
4.4 1.15524992346764
3.6 1.15524992346764
3.6 1.02589786052704
};
\addplot [ultra thick, steelblue31119180]
table {%
4 1.02589786052704
4 0.916501820087433
};
\addplot [ultra thick, steelblue31119180]
table {%
4 1.15524992346764
4 1.25098788738251
};
\addplot [ultra thick, steelblue31119180]
table {%
3.8 0.916501820087433
4.2 0.916501820087433
};
\addplot [ultra thick, steelblue31119180]
table {%
3.8 1.25098788738251
4.2 1.25098788738251
};
\addplot [ultra thick, steelblue31119180]
table {%
4.6 0.671762555837631
5.4 0.671762555837631
5.4 0.851939216256142
4.6 0.851939216256142
4.6 0.671762555837631
};
\addplot [ultra thick, steelblue31119180]
table {%
5 0.671762555837631
5 0.584396123886108
};
\addplot [ultra thick, steelblue31119180]
table {%
5 0.851939216256142
5 1.04319095611572
};
\addplot [ultra thick, steelblue31119180]
table {%
4.8 0.584396123886108
5.2 0.584396123886108
};
\addplot [ultra thick, steelblue31119180]
table {%
4.8 1.04319095611572
5.2 1.04319095611572
};
\addplot [ultra thick, steelblue31119180]
table {%
-0.4 0.709299594163895
0.4 0.709299594163895
};
\addplot [forestgreen4416044, mark=triangle*, mark size=3, mark options={solid}, only marks]
table {%
0 0.741395997149604
};
\addplot [ultra thick, steelblue31119180]
table {%
0.6 0.546452432870865
1.4 0.546452432870865
};
\addplot [forestgreen4416044, mark=triangle*, mark size=3, mark options={solid}, only marks]
table {%
1 0.558016712112086
};
\addplot [ultra thick, steelblue31119180]
table {%
1.6 0.458221465349197
2.4 0.458221465349197
};
\addplot [forestgreen4416044, mark=triangle*, mark size=3, mark options={solid}, only marks]
table {%
2 0.462062742028918
};
\addplot [ultra thick, steelblue31119180]
table {%
2.6 0.503057435154915
3.4 0.503057435154915
};
\addplot [forestgreen4416044, mark=triangle*, mark size=3, mark options={solid}, only marks]
table {%
3 0.535144391868796
};
\addplot [ultra thick, steelblue31119180]
table {%
3.6 1.08842092752457
4.4 1.08842092752457
};
\addplot [forestgreen4416044, mark=triangle*, mark size=3, mark options={solid}, only marks]
table {%
4 1.08559619528907
};
\addplot [ultra thick, steelblue31119180]
table {%
4.6 0.742198139429092
5.4 0.742198139429092
};
\addplot [forestgreen4416044, mark=triangle*, mark size=3, mark options={solid}, only marks]
table {%
5 0.774040730936187
};
\end{axis}

\end{tikzpicture}

%% file: tikz/exp2_scoreq.tex
% This file was created with tikzplotlib v0.10.1.
\begin{tikzpicture}

\definecolor{darkgray176}{RGB}{176,176,176}
\definecolor{forestgreen4416044}{RGB}{44,160,44}
\definecolor{indianred1967882}{RGB}{196,78,82}
\definecolor{steelblue76114176}{RGB}{76,114,176}

\begin{axis}[
tick align=outside,
tick pos=left,
x grid style={darkgray176},
xmajorgrids,
xmin=-0.5, xmax=6.5,
xlabel={},
xticklabels={},
y grid style={darkgray176},
ymajorgrids,
ylabel={},
yticklabels={},
ymin=0.10, ymax=1.3,
ytick style={color=black},
]

% add vertical line to separate retrained models from pretrained ones
\addplot[black, dashed, thick, domain=-1:10] coordinates {(3.5,-1) (3.5,10)};;

\addplot [ultra thick, steelblue76114176]
table {%
-0.4 0.387603476643562
0.4 0.387603476643562
0.4 0.51121823489666
-0.4 0.51121823489666
-0.4 0.387603476643562
};
\addplot [ultra thick, steelblue76114176]
table {%
0 0.387603476643562
0 0.301101207733154
};
\addplot [ultra thick, steelblue76114176]
table {%
0 0.51121823489666
0 0.692492306232452
};
\addplot [ultra thick, steelblue76114176]
table {%
-0.2 0.301101207733154
0.2 0.301101207733154
};
\addplot [ultra thick, steelblue76114176]
table {%
-0.2 0.692492306232452
0.2 0.692492306232452
};
\addplot [black, mark=o, mark size=3, mark options={solid,fill opacity=0,draw=steelblue76114176}, only marks]
table {%
0 0.726335167884827
};
\addplot [ultra thick, steelblue76114176]
table {%
0.6 0.212978702038527
1.4 0.212978702038527
1.4 0.283236153423786
0.6 0.283236153423786
0.6 0.212978702038527
};
\addplot [ultra thick, steelblue76114176]
table {%
1 0.212978702038527
1 0.159924566745758
};
\addplot [ultra thick, steelblue76114176]
table {%
1 0.283236153423786
1 0.382632434368133
};
\addplot [ultra thick, steelblue76114176]
table {%
0.8 0.159924566745758
1.2 0.159924566745758
};
\addplot [ultra thick, steelblue76114176]
table {%
0.8 0.382632434368133
1.2 0.382632434368133
};
\addplot [black, mark=o, mark size=3, mark options={solid,fill opacity=0,draw=steelblue76114176}, only marks]
table {%
1 0.398969799280167
};
\addplot [ultra thick, indianred1967882]
table {%
1.6 0.266204938292503
2.4 0.266204938292503
2.4 0.414339311420918
1.6 0.414339311420918
1.6 0.266204938292503
};
\addplot [ultra thick, indianred1967882]
table {%
2 0.266204938292503
2 0.230220511555672
};
\addplot [ultra thick, indianred1967882]
table {%
2 0.414339311420918
2 0.541127920150757
};
\addplot [ultra thick, indianred1967882]
table {%
1.8 0.230220511555672
2.2 0.230220511555672
};
\addplot [ultra thick, indianred1967882]
table {%
1.8 0.541127920150757
2.2 0.541127920150757
};
\addplot [ultra thick, indianred1967882]
table {%
2.6 0.312813468277454
3.4 0.312813468277454
3.4 0.462297074496746
2.6 0.462297074496746
2.6 0.312813468277454
};
\addplot [ultra thick, indianred1967882]
table {%
3 0.312813468277454
3 0.196022853255272
};
\addplot [ultra thick, indianred1967882]
table {%
3 0.462297074496746
3 0.675450921058655
};
\addplot [ultra thick, indianred1967882]
table {%
2.8 0.196022853255272
3.2 0.196022853255272
};
\addplot [ultra thick, indianred1967882]
table {%
2.8 0.675450921058655
3.2 0.675450921058655
};
\addplot [ultra thick, indianred1967882]
table {%
3.6 0.799071654677391
4.4 0.799071654677391
4.4 0.946784675121307
3.6 0.946784675121307
3.6 0.799071654677391
};
\addplot [ultra thick, indianred1967882]
table {%
4 0.799071654677391
4 0.683956205844879
};
\addplot [ultra thick, indianred1967882]
table {%
4 0.946784675121307
4 1.04927802085876
};
\addplot [ultra thick, indianred1967882]
table {%
3.8 0.683956205844879
4.2 0.683956205844879
};
\addplot [ultra thick, indianred1967882]
table {%
3.8 1.04927802085876
4.2 1.04927802085876
};
\addplot [ultra thick, steelblue76114176]
table {%
4.6 0.807793378829956
5.4 0.807793378829956
5.4 0.949133694171905
4.6 0.949133694171905
4.6 0.807793378829956
};
\addplot [ultra thick, steelblue76114176]
table {%
5 0.807793378829956
5 0.653511047363281
};
\addplot [ultra thick, steelblue76114176]
table {%
5 0.949133694171905
5 0.985864102840424
};
\addplot [ultra thick, steelblue76114176]
table {%
4.8 0.653511047363281
5.2 0.653511047363281
};
\addplot [ultra thick, steelblue76114176]
table {%
4.8 0.985864102840424
5.2 0.985864102840424
};
\addplot [ultra thick, steelblue76114176]
table {%
5.6 0.905823022127152
6.4 0.905823022127152
6.4 1.03468656539917
5.6 1.03468656539917
5.6 0.905823022127152
};
\addplot [ultra thick, steelblue76114176]
table {%
6 0.905823022127152
6 0.79025787115097
};
\addplot [ultra thick, steelblue76114176]
table {%
6 1.03468656539917
6 1.10464787483215
};
\addplot [ultra thick, steelblue76114176]
table {%
5.8 0.79025787115097
6.2 0.79025787115097
};
\addplot [ultra thick, steelblue76114176]
table {%
5.8 1.10464787483215
6.2 1.10464787483215
};
\addplot [black, mark=o, mark size=3, mark options={solid,fill opacity=0,draw=steelblue76114176}, only marks]
table {%
6 0.697642624378204
};
\addplot [ultra thick, steelblue76114176]
table {%
-0.4 0.458221465349197
0.4 0.458221465349197
};
\addplot [forestgreen4416044, mark=triangle*, mark size=3, mark options={solid}, only marks]
table {%
0 0.462062742028918
};
\addplot [ultra thick, steelblue76114176]
table {%
0.6 0.235234305262566
1.4 0.235234305262566
};
\addplot [forestgreen4416044, mark=triangle*, mark size=3, mark options={solid}, only marks]
table {%
1 0.256052782492978
};
\addplot [ultra thick, indianred1967882]
table {%
1.6 0.331205502152443
2.4 0.331205502152443
};
\addplot [forestgreen4416044, mark=triangle*, mark size=3, mark options={solid}, only marks]
table {%
2 0.345831625695739
};
\addplot [ultra thick, indianred1967882]
table {%
2.6 0.346161976456642
3.4 0.346161976456642
};
\addplot [forestgreen4416044, mark=triangle*, mark size=3, mark options={solid}, only marks]
table {%
3 0.379830745181867
};
\addplot [ultra thick, indianred1967882]
table {%
3.6 0.869785338640213
4.4 0.869785338640213
};
\addplot [forestgreen4416044, mark=triangle*, mark size=3, mark options={solid}, only marks]
table {%
4 0.86789558827877
};
\addplot [ultra thick, steelblue76114176]
table {%
4.6 0.89818674325943
5.4 0.89818674325943
};
\addplot [forestgreen4416044, mark=triangle*, mark size=3, mark options={solid}, only marks]
table {%
5 0.872002382363592
};
\addplot [ultra thick, steelblue76114176]
table {%
5.6 0.996009290218353
6.4 0.996009290218353
};
\addplot [forestgreen4416044, mark=triangle*, mark size=3, mark options={solid}, only marks]
table {%
6 0.973079617534365
};
\end{axis}

\end{tikzpicture}

%% file: tikz/exp3_scoreq.tex
% This file was created with tikzplotlib v0.10.1.
\begin{tikzpicture}

\definecolor{darkgray141160203}{RGB}{141,160,203}
\definecolor{darkgray176}{RGB}{176,176,176}
\definecolor{forestgreen4416044}{RGB}{44,160,44}
\definecolor{lightgray204}{RGB}{204,204,204}
\definecolor{mediumaquamarine102194165}{RGB}{102,194,165}
\definecolor{salmon25214198}{RGB}{252,141,98}

\begin{axis}[
tick align=outside,
tick pos=left,
x grid style={darkgray176},
xmajorgrids,
xmin=-0.5, xmax=3.5,
xlabel={},
xticklabels={},
y grid style={darkgray176},
ymajorgrids,
ylabel={},
yticklabels={},
ymin=0.10,
ymax=1.3,
ytick style={color=black},
legend style={at={(0.5, 0.7)}, anchor=south},
]

%\draw[draw=mediumaquamarine102194165,ultra thick] (axis cs:0,0) rectangle (axis cs:0,0);
\addlegendimage{draw=mediumaquamarine102194165,ultra thick}
\addlegendentry{1.5 kbps}

%\draw[draw=salmon25214198,ultra thick] (axis cs:0,0) rectangle (axis cs:0,0);
\addlegendimage{draw=salmon25214198,ultra thick}
\addlegendentry{3.0 kbps}

%\draw[draw=darkgray141160203,ultra thick] (axis cs:0,0) rectangle (axis cs:0,0);
\addlegendimage{draw=darkgray141160203,ultra thick}
\addlegendentry{6.0 kbps}

\addplot [ultra thick, mediumaquamarine102194165, forget plot]
table {%
-0.4 0.469410449266434
-0.133333333333333 0.469410449266434
-0.133333333333333 0.59988659620285
-0.4 0.59988659620285
-0.4 0.469410449266434
};
\addplot [ultra thick, mediumaquamarine102194165, forget plot]
table {%
-0.266666666666667 0.469410449266434
-0.266666666666667 0.362280517816544
};
\addplot [ultra thick, mediumaquamarine102194165, forget plot]
table {%
-0.266666666666667 0.59988659620285
-0.266666666666667 0.732109010219574
};
\addplot [ultra thick, mediumaquamarine102194165, forget plot]
table {%
-0.333333333333333 0.362280517816544
-0.2 0.362280517816544
};
\addplot [ultra thick, mediumaquamarine102194165, forget plot]
table {%
-0.333333333333333 0.732109010219574
-0.2 0.732109010219574
};
\addplot [black, mark=o, mark size=3, mark options={solid,fill opacity=0,draw=mediumaquamarine102194165}, only marks, forget plot]
table {%
-0.266666666666667 1.01817452907562
};
\addplot [ultra thick, mediumaquamarine102194165, forget plot]
table {%
0.6 0.280393347144127
0.866666666666667 0.280393347144127
0.866666666666667 0.381597846746445
0.6 0.381597846746445
0.6 0.280393347144127
};
\addplot [ultra thick, mediumaquamarine102194165, forget plot]
table {%
0.733333333333333 0.280393347144127
0.733333333333333 0.242222562432289
};
\addplot [ultra thick, mediumaquamarine102194165, forget plot]
table {%
0.733333333333333 0.381597846746445
0.733333333333333 0.513123214244843
};
\addplot [ultra thick, mediumaquamarine102194165, forget plot]
table {%
0.666666666666667 0.242222562432289
0.8 0.242222562432289
};
\addplot [ultra thick, mediumaquamarine102194165, forget plot]
table {%
0.666666666666667 0.513123214244843
0.8 0.513123214244843
};
\addplot [black, mark=o, mark size=3, mark options={solid,fill opacity=0,draw=mediumaquamarine102194165}, only marks, forget plot]
table {%
0.733333333333333 0.634301543235779
0.733333333333333 0.761786639690399
};
\addplot [ultra thick, mediumaquamarine102194165, forget plot]
table {%
1.6 0.284306868910789
1.86666666666667 0.284306868910789
1.86666666666667 0.333586052060127
1.6 0.333586052060127
1.6 0.284306868910789
};
\addplot [ultra thick, mediumaquamarine102194165, forget plot]
table {%
1.73333333333333 0.284306868910789
1.73333333333333 0.215377554297447
};
\addplot [ultra thick, mediumaquamarine102194165, forget plot]
table {%
1.73333333333333 0.333586052060127
1.73333333333333 0.392172992229462
};
\addplot [ultra thick, mediumaquamarine102194165, forget plot]
table {%
1.66666666666667 0.215377554297447
1.8 0.215377554297447
};
\addplot [ultra thick, mediumaquamarine102194165, forget plot]
table {%
1.66666666666667 0.392172992229462
1.8 0.392172992229462
};
\addplot [black, mark=o, mark size=3, mark options={solid,fill opacity=0,draw=mediumaquamarine102194165}, only marks, forget plot]
table {%
1.73333333333333 0.456760168075562
};
\addplot [ultra thick, mediumaquamarine102194165, forget plot]
table {%
2.6 0.498331442475319
2.86666666666667 0.498331442475319
2.86666666666667 0.693927302956581
2.6 0.693927302956581
2.6 0.498331442475319
};
\addplot [ultra thick, mediumaquamarine102194165, forget plot]
table {%
2.73333333333333 0.498331442475319
2.73333333333333 0.344816088676453
};
\addplot [ultra thick, mediumaquamarine102194165, forget plot]
table {%
2.73333333333333 0.693927302956581
2.73333333333333 0.909660339355469
};
\addplot [ultra thick, mediumaquamarine102194165, forget plot]
table {%
2.66666666666667 0.344816088676453
2.8 0.344816088676453
};
\addplot [ultra thick, mediumaquamarine102194165, forget plot]
table {%
2.66666666666667 0.909660339355469
2.8 0.909660339355469
};
\addplot [black, mark=o, mark size=3, mark options={solid,fill opacity=0,draw=mediumaquamarine102194165}, only marks, forget plot]
table {%
2.73333333333333 1.0230747461319
};
\addplot [ultra thick, salmon25214198, forget plot]
table {%
-0.133333333333333 0.387603476643562
0.133333333333333 0.387603476643562
0.133333333333333 0.51121823489666
-0.133333333333333 0.51121823489666
-0.133333333333333 0.387603476643562
};
\addplot [ultra thick, salmon25214198, forget plot]
table {%
0 0.387603476643562
0 0.301101207733154
};
\addplot [ultra thick, salmon25214198, forget plot]
table {%
0 0.51121823489666
0 0.692492306232452
};
\addplot [ultra thick, salmon25214198, forget plot]
table {%
-0.0666666666666667 0.301101207733154
0.0666666666666667 0.301101207733154
};
\addplot [ultra thick, salmon25214198, forget plot]
table {%
-0.0666666666666667 0.692492306232452
0.0666666666666667 0.692492306232452
};
\addplot [black, mark=o, mark size=3, mark options={solid,fill opacity=0,draw=salmon25214198}, only marks, forget plot]
table {%
0 0.726335167884827
};
\addplot [ultra thick, salmon25214198, forget plot]
table {%
0.866666666666667 0.212978702038527
1.13333333333333 0.212978702038527
1.13333333333333 0.283236153423786
0.866666666666667 0.283236153423786
0.866666666666667 0.212978702038527
};
\addplot [ultra thick, salmon25214198, forget plot]
table {%
1 0.212978702038527
1 0.159924566745758
};
\addplot [ultra thick, salmon25214198, forget plot]
table {%
1 0.283236153423786
1 0.382632434368133
};
\addplot [ultra thick, salmon25214198, forget plot]
table {%
0.933333333333333 0.159924566745758
1.06666666666667 0.159924566745758
};
\addplot [ultra thick, salmon25214198, forget plot]
table {%
0.933333333333333 0.382632434368133
1.06666666666667 0.382632434368133
};
\addplot [black, mark=o, mark size=3, mark options={solid,fill opacity=0,draw=salmon25214198}, only marks, forget plot]
table {%
1 0.398969799280167
};
\addplot [ultra thick, salmon25214198, forget plot]
table {%
1.86666666666667 0.230350218713284
2.13333333333333 0.230350218713284
2.13333333333333 0.348886884748936
1.86666666666667 0.348886884748936
1.86666666666667 0.230350218713284
};
\addplot [ultra thick, salmon25214198, forget plot]
table {%
2 0.230350218713284
2 0.198185905814171
};
\addplot [ultra thick, salmon25214198, forget plot]
table {%
2 0.348886884748936
2 0.428218394517899
};
\addplot [ultra thick, salmon25214198, forget plot]
table {%
1.93333333333333 0.198185905814171
2.06666666666667 0.198185905814171
};
\addplot [ultra thick, salmon25214198, forget plot]
table {%
1.93333333333333 0.428218394517899
2.06666666666667 0.428218394517899
};
\addplot [ultra thick, salmon25214198, forget plot]
table {%
2.86666666666667 0.312813468277454
3.13333333333333 0.312813468277454
3.13333333333333 0.462297074496746
2.86666666666667 0.462297074496746
2.86666666666667 0.312813468277454
};
\addplot [ultra thick, salmon25214198, forget plot]
table {%
3 0.312813468277454
3 0.196022853255272
};
\addplot [ultra thick, salmon25214198, forget plot]
table {%
3 0.462297074496746
3 0.675450921058655
};
\addplot [ultra thick, salmon25214198, forget plot]
table {%
2.93333333333333 0.196022853255272
3.06666666666667 0.196022853255272
};
\addplot [ultra thick, salmon25214198, forget plot]
table {%
2.93333333333333 0.675450921058655
3.06666666666667 0.675450921058655
};
\addplot [ultra thick, darkgray141160203, forget plot]
table {%
0.133333333333333 0.267101816833019
0.4 0.267101816833019
0.4 0.324909783899784
0.133333333333333 0.324909783899784
0.133333333333333 0.267101816833019
};
\addplot [ultra thick, darkgray141160203, forget plot]
table {%
0.266666666666667 0.267101816833019
0.266666666666667 0.229920834302902
};
\addplot [ultra thick, darkgray141160203, forget plot]
table {%
0.266666666666667 0.324909783899784
0.266666666666667 0.380650669336319
};
\addplot [ultra thick, darkgray141160203, forget plot]
table {%
0.2 0.229920834302902
0.333333333333333 0.229920834302902
};
\addplot [ultra thick, darkgray141160203, forget plot]
table {%
0.2 0.380650669336319
0.333333333333333 0.380650669336319
};
\addplot [black, mark=o, mark size=3, mark options={solid,fill opacity=0,draw=darkgray141160203}, only marks, forget plot]
table {%
0.266666666666667 0.518310964107513
0.266666666666667 0.495242267847061
0.266666666666667 0.414562195539474
};
\addplot [ultra thick, darkgray141160203, forget plot]
table {%
1.13333333333333 0.158590756356716
1.4 0.158590756356716
1.4 0.204425312578678
1.13333333333333 0.204425312578678
1.13333333333333 0.158590756356716
};
\addplot [ultra thick, darkgray141160203, forget plot]
table {%
1.26666666666667 0.158590756356716
1.26666666666667 0.136654183268547
};
\addplot [ultra thick, darkgray141160203, forget plot]
table {%
1.26666666666667 0.204425312578678
1.26666666666667 0.265159636735916
};
\addplot [ultra thick, darkgray141160203, forget plot]
table {%
1.2 0.136654183268547
1.33333333333333 0.136654183268547
};
\addplot [ultra thick, darkgray141160203, forget plot]
table {%
1.2 0.265159636735916
1.33333333333333 0.265159636735916
};
\addplot [black, mark=o, mark size=3, mark options={solid,fill opacity=0,draw=darkgray141160203}, only marks, forget plot]
table {%
1.26666666666667 0.341052621603012
};
\addplot [ultra thick, darkgray141160203, forget plot]
table {%
2.13333333333333 0.177329894155264
2.4 0.177329894155264
2.4 0.229287482798099
2.13333333333333 0.229287482798099
2.13333333333333 0.177329894155264
};
\addplot [ultra thick, darkgray141160203, forget plot]
table {%
2.26666666666667 0.177329894155264
2.26666666666667 0.150958105921745
};
\addplot [ultra thick, darkgray141160203, forget plot]
table {%
2.26666666666667 0.229287482798099
2.26666666666667 0.268011480569839
};
\addplot [ultra thick, darkgray141160203, forget plot]
table {%
2.2 0.150958105921745
2.33333333333333 0.150958105921745
};
\addplot [ultra thick, darkgray141160203, forget plot]
table {%
2.2 0.268011480569839
2.33333333333333 0.268011480569839
};
\addplot [black, mark=o, mark size=3, mark options={solid,fill opacity=0,draw=darkgray141160203}, only marks, forget plot]
table {%
2.26666666666667 0.308078467845917
2.26666666666667 0.338074922561646
};
\addplot [ultra thick, darkgray141160203, forget plot]
table {%
3.13333333333333 0.206100020557642
3.4 0.206100020557642
3.4 0.301869429647923
3.13333333333333 0.301869429647923
3.13333333333333 0.206100020557642
};
\addplot [ultra thick, darkgray141160203, forget plot]
table {%
3.26666666666667 0.206100020557642
3.26666666666667 0.14718647301197
};
\addplot [ultra thick, darkgray141160203, forget plot]
table {%
3.26666666666667 0.301869429647923
3.26666666666667 0.396000981330872
};
\addplot [ultra thick, darkgray141160203, forget plot]
table {%
3.2 0.14718647301197
3.33333333333333 0.14718647301197
};
\addplot [ultra thick, darkgray141160203, forget plot]
table {%
3.2 0.396000981330872
3.33333333333333 0.396000981330872
};
\addplot [ultra thick, mediumaquamarine102194165, forget plot]
table {%
-0.4 0.511586487293243
-0.133333333333333 0.511586487293243
};
\addplot [forestgreen4416044, mark=triangle*, mark size=3, mark options={solid}, only marks, forget plot]
table {%
-0.266666666666667 0.541036769747734
};
\addplot [ultra thick, mediumaquamarine102194165, forget plot]
table {%
0.6 0.320612877607346
0.866666666666667 0.320612877607346
};
\addplot [forestgreen4416044, mark=triangle*, mark size=3, mark options={solid}, only marks, forget plot]
table {%
0.733333333333333 0.358196255884
};
\addplot [ultra thick, mediumaquamarine102194165, forget plot]
table {%
1.6 0.311950474977493
1.86666666666667 0.311950474977493
};
\addplot [forestgreen4416044, mark=triangle*, mark size=3, mark options={solid}, only marks, forget plot]
table {%
1.73333333333333 0.313876253153597
};
\addplot [ultra thick, mediumaquamarine102194165, forget plot]
table {%
2.6 0.590329498052597
2.86666666666667 0.590329498052597
};
\addplot [forestgreen4416044, mark=triangle*, mark size=3, mark options={solid}, only marks, forget plot]
table {%
2.73333333333333 0.61259162319558
};
\addplot [ultra thick, salmon25214198, forget plot]
table {%
-0.133333333333333 0.458221465349197
0.133333333333333 0.458221465349197
};
\addplot [forestgreen4416044, mark=triangle*, mark size=3, mark options={solid}, only marks, forget plot]
table {%
0 0.462062742028918
};
\addplot [ultra thick, salmon25214198, forget plot]
table {%
0.866666666666667 0.235234305262566
1.13333333333333 0.235234305262566
};
\addplot [forestgreen4416044, mark=triangle*, mark size=3, mark options={solid}, only marks, forget plot]
table {%
1 0.256052782492978
};
\addplot [ultra thick, salmon25214198, forget plot]
table {%
1.86666666666667 0.276035860180855
2.13333333333333 0.276035860180855
};
\addplot [forestgreen4416044, mark=triangle*, mark size=3, mark options={solid}, only marks, forget plot]
table {%
2 0.291672651256834
};
\addplot [ultra thick, salmon25214198, forget plot]
table {%
2.86666666666667 0.346161976456642
3.13333333333333 0.346161976456642
};
\addplot [forestgreen4416044, mark=triangle*, mark size=3, mark options={solid}, only marks, forget plot]
table {%
3 0.379830745181867
};
\addplot [ultra thick, darkgray141160203, forget plot]
table {%
0.133333333333333 0.285816997289658
0.4 0.285816997289658
};
\addplot [forestgreen4416044, mark=triangle*, mark size=3, mark options={solid}, only marks, forget plot]
table {%
0.266666666666667 0.304519809782505
};
\addplot [ultra thick, darkgray141160203, forget plot]
table {%
1.13333333333333 0.171802401542664
1.4 0.171802401542664
};
\addplot [forestgreen4416044, mark=triangle*, mark size=3, mark options={solid}, only marks, forget plot]
table {%
1.26666666666667 0.188425527619464
};
\addplot [ultra thick, darkgray141160203, forget plot]
table {%
2.13333333333333 0.200180530548096
2.4 0.200180530548096
};
\addplot [forestgreen4416044, mark=triangle*, mark size=3, mark options={solid}, only marks, forget plot]
table {%
2.26666666666667 0.208694283451353
};
\addplot [ultra thick, darkgray141160203, forget plot]
table {%
3.13333333333333 0.236896641552448
3.4 0.236896641552448
};
\addplot [forestgreen4416044, mark=triangle*, mark size=3, mark options={solid}, only marks, forget plot]
table {%
3.26666666666667 0.257894038621868
};
\end{axis}

\end{tikzpicture}

%% file: tikz/exp1_visqol.tex
% This file was created with tikzplotlib v0.10.1.
\begin{tikzpicture}

\definecolor{darkgray176}{RGB}{176,176,176}
\definecolor{forestgreen4416044}{RGB}{44,160,44}
\definecolor{steelblue31119180}{RGB}{31,119,180}

\begin{axis}[
tick align=outside,
tick pos=left,
x grid style={darkgray176},
xmin=-0.5, xmax=5.5,
xtick style={color=black},
xticklabel style={rotate=90.0, font=\plotTickFont},
xticklabels={,\mTw{}, \lTw{}, \mTm{}, \lTm{}, \mTl{}, \lTl{}},
xmajorgrids,
y grid style={darkgray176},
ylabel={$\longrightarrow$ ~ \visqol{}},
ylabel style={yshift=-10pt},
ymajorgrids,
ymin=1.72, ymax=4.0,
ytick style={color=black},
clip=true
]
\addplot [ultra thick, steelblue31119180]
table {%
-0.4 2.2822875
0.4 2.2822875
0.4 2.4888175
-0.4 2.4888175
-0.4 2.2822875
};
\addplot [ultra thick, steelblue31119180]
table {%
0 2.2822875
0 2.22211
};
\addplot [ultra thick, steelblue31119180]
table {%
0 2.4888175
0 2.66178
};
\addplot [ultra thick, steelblue31119180]
table {%
-0.2 2.22211
0.2 2.22211
};
\addplot [ultra thick, steelblue31119180]
table {%
-0.2 2.66178
0.2 2.66178
};
\addplot [ultra thick, steelblue31119180]
table {%
0.6 2.583745
1.4 2.583745
1.4 2.729065
0.6 2.729065
0.6 2.583745
};
\addplot [ultra thick, steelblue31119180]
table {%
1 2.583745
1 2.49329
};
\addplot [ultra thick, steelblue31119180]
table {%
1 2.729065
1 2.91149
};
\addplot [ultra thick, steelblue31119180]
table {%
0.8 2.49329
1.2 2.49329
};
\addplot [ultra thick, steelblue31119180]
table {%
0.8 2.91149
1.2 2.91149
};
\addplot [black, mark=o, mark size=3, mark options={solid,fill opacity=0,draw=steelblue31119180}, only marks]
table {%
1 2.28228
1 3.09111
};
\addplot [ultra thick, steelblue31119180]
table {%
1.6 2.82464
2.4 2.82464
2.4 3.0300575
1.6 3.0300575
1.6 2.82464
};
\addplot [ultra thick, steelblue31119180]
table {%
2 2.82464
2 2.54078
};
\addplot [ultra thick, steelblue31119180]
table {%
2 3.0300575
2 3.20434
};
\addplot [ultra thick, steelblue31119180]
table {%
1.8 2.54078
2.2 2.54078
};
\addplot [ultra thick, steelblue31119180]
table {%
1.8 3.20434
2.2 3.20434
};
\addplot [ultra thick, steelblue31119180]
table {%
2.6 2.56449
3.4 2.56449
3.4 2.6879625
2.6 2.6879625
2.6 2.56449
};
\addplot [ultra thick, steelblue31119180]
table {%
3 2.56449
3 2.40186
};
\addplot [ultra thick, steelblue31119180]
table {%
3 2.6879625
3 2.81961
};
\addplot [ultra thick, steelblue31119180]
table {%
2.8 2.40186
3.2 2.40186
};
\addplot [ultra thick, steelblue31119180]
table {%
2.8 2.81961
3.2 2.81961
};
\addplot [black, mark=o, mark size=3, mark options={solid,fill opacity=0,draw=steelblue31119180}, only marks]
table {%
3 2.8776
};
\addplot [ultra thick, steelblue31119180]
table {%
3.6 2.17164
4.4 2.17164
4.4 2.3743325
3.6 2.3743325
3.6 2.17164
};
\addplot [ultra thick, steelblue31119180]
table {%
4 2.17164
4 2.07234
};
\addplot [ultra thick, steelblue31119180]
table {%
4 2.3743325
4 2.55802
};
\addplot [ultra thick, steelblue31119180]
table {%
3.8 2.07234
4.2 2.07234
};
\addplot [ultra thick, steelblue31119180]
table {%
3.8 2.55802
4.2 2.55802
};
\addplot [ultra thick, steelblue31119180]
table {%
4.6 2.49901
5.4 2.49901
5.4 2.7052625
4.6 2.7052625
4.6 2.49901
};
\addplot [ultra thick, steelblue31119180]
table {%
5 2.49901
5 2.42468
};
\addplot [ultra thick, steelblue31119180]
table {%
5 2.7052625
5 2.85094
};
\addplot [ultra thick, steelblue31119180]
table {%
4.8 2.42468
5.2 2.42468
};
\addplot [ultra thick, steelblue31119180]
table {%
4.8 2.85094
5.2 2.85094
};
\addplot [black, mark=o, mark size=3, mark options={solid,fill opacity=0,draw=steelblue31119180}, only marks]
table {%
5 3.04385
};
\addplot [ultra thick, steelblue31119180]
table {%
-0.4 2.3958
0.4 2.3958
};
\addplot [forestgreen4416044, mark=triangle*, mark size=3, mark options={solid}, only marks]
table {%
0 2.392265
};
\addplot [ultra thick, steelblue31119180]
table {%
0.6 2.65192
1.4 2.65192
};
\addplot [forestgreen4416044, mark=triangle*, mark size=3, mark options={solid}, only marks]
table {%
1 2.65883392857143
};
\addplot [ultra thick, steelblue31119180]
table {%
1.6 2.90149
2.4 2.90149
};
\addplot [forestgreen4416044, mark=triangle*, mark size=3, mark options={solid}, only marks]
table {%
2 2.92690892857143
};
\addplot [ultra thick, steelblue31119180]
table {%
2.6 2.625995
3.4 2.625995
};
\addplot [forestgreen4416044, mark=triangle*, mark size=3, mark options={solid}, only marks]
table {%
3 2.6182675
};
\addplot [ultra thick, steelblue31119180]
table {%
3.6 2.261475
4.4 2.261475
};
\addplot [forestgreen4416044, mark=triangle*, mark size=3, mark options={solid}, only marks]
table {%
4 2.28943535714286
};
\addplot [ultra thick, steelblue31119180]
table {%
4.6 2.60314
5.4 2.60314
};
\addplot [forestgreen4416044, mark=triangle*, mark size=3, mark options={solid}, only marks]
table {%
5 2.62179714285714
};
\end{axis}

\end{tikzpicture}

%% file: tikz/exp2_visqol.tex
% This file was created with tikzplotlib v0.10.1.
\begin{tikzpicture}

\definecolor{darkgray176}{RGB}{176,176,176}
\definecolor{forestgreen4416044}{RGB}{44,160,44}
\definecolor{indianred1967882}{RGB}{196,78,82}
\definecolor{steelblue76114176}{RGB}{76,114,176}

\begin{axis}[
tick align=outside,
tick pos=left,
x grid style={darkgray176},
xmajorgrids,
xmin=-0.5, xmax=6.5,
xtick style={color=black},
xticklabel style={rotate=90.0, font=\plotTickFont},
xticklabels={, \mTm{}, \shortstack{\mTm{}\\fine-tuned}, {retr. \EC{}}, \QGAN{}, {pretr. \EC{}}, \MBD{}, \LDC{}},
y grid style={darkgray176},
ymajorgrids,
ylabel={},
yticklabels={},
ymin=1.72, ymax=4.0,
ytick style={color=black},
]

% add vertical line to separate retrained models from pretrained ones
\addplot[black, dashed, thick, domain=-1:10] coordinates {(3.5,-1) (3.5,10)};;

\addplot [ultra thick, steelblue76114176]
table {%
-0.4 2.82464
0.4 2.82464
0.4 3.0300575
-0.4 3.0300575
-0.4 2.82464
};
\addplot [ultra thick, steelblue76114176]
table {%
0 2.82464
0 2.54078
};
\addplot [ultra thick, steelblue76114176]
table {%
0 3.0300575
0 3.20434
};
\addplot [ultra thick, steelblue76114176]
table {%
-0.2 2.54078
0.2 2.54078
};
\addplot [ultra thick, steelblue76114176]
table {%
-0.2 3.20434
0.2 3.20434
};
\addplot [ultra thick, steelblue76114176]
table {%
0.6 2.9338625
1.4 2.9338625
1.4 3.0295975
0.6 3.0295975
0.6 2.9338625
};
\addplot [ultra thick, steelblue76114176]
table {%
1 2.9338625
1 2.85716
};
\addplot [ultra thick, steelblue76114176]
table {%
1 3.0295975
1 3.15033
};
\addplot [ultra thick, steelblue76114176]
table {%
0.8 2.85716
1.2 2.85716
};
\addplot [ultra thick, steelblue76114176]
table {%
0.8 3.15033
1.2 3.15033
};
\addplot [black, mark=o, mark size=3, mark options={solid,fill opacity=0,draw=steelblue76114176}, only marks]
table {%
1 2.76701
1 3.37257
1 3.25144
1 3.21641
};
\addplot [ultra thick, indianred1967882]
table {%
1.6 2.987895
2.4 2.987895
2.4 3.18613
1.6 3.18613
1.6 2.987895
};
\addplot [ultra thick, indianred1967882]
table {%
2 2.987895
2 2.87927
};
\addplot [ultra thick, indianred1967882]
table {%
2 3.18613
2 3.41908
};
\addplot [ultra thick, indianred1967882]
table {%
1.8 2.87927
2.2 2.87927
};
\addplot [ultra thick, indianred1967882]
table {%
1.8 3.41908
2.2 3.41908
};
\addplot [black, mark=o, mark size=3, mark options={solid,fill opacity=0,draw=indianred1967882}, only marks]
table {%
2 3.57374
};
\addplot [ultra thick, indianred1967882]
table {%
2.6 2.78094
3.4 2.78094
3.4 2.9932775
2.6 2.9932775
2.6 2.78094
};
\addplot [ultra thick, indianred1967882]
table {%
3 2.78094
3 2.57863
};
\addplot [ultra thick, indianred1967882]
table {%
3 2.9932775
3 3.20405
};
\addplot [ultra thick, indianred1967882]
table {%
2.8 2.57863
3.2 2.57863
};
\addplot [ultra thick, indianred1967882]
table {%
2.8 3.20405
3.2 3.20405
};
\addplot [ultra thick, indianred1967882]
table {%
3.6 2.9224325
4.4 2.9224325
4.4 3.0985625
3.6 3.0985625
3.6 2.9224325
};
\addplot [ultra thick, indianred1967882]
table {%
4 2.9224325
4 2.69482
};
\addplot [ultra thick, indianred1967882]
table {%
4 3.0985625
4 3.28616
};
\addplot [ultra thick, indianred1967882]
table {%
3.8 2.69482
4.2 2.69482
};
\addplot [ultra thick, indianred1967882]
table {%
3.8 3.28616
4.2 3.28616
};
\addplot [ultra thick, steelblue76114176]
table {%
4.6 2.3414475
5.4 2.3414475
5.4 2.5377275
4.6 2.5377275
4.6 2.3414475
};
\addplot [ultra thick, steelblue76114176]
table {%
5 2.3414475
5 2.14835
};
\addplot [ultra thick, steelblue76114176]
table {%
5 2.5377275
5 2.78799
};
\addplot [ultra thick, steelblue76114176]
table {%
4.8 2.14835
5.2 2.14835
};
\addplot [ultra thick, steelblue76114176]
table {%
4.8 2.78799
5.2 2.78799
};
\addplot [ultra thick, steelblue76114176]
table {%
5.6 2.0287
6.4 2.0287
6.4 2.20253
5.6 2.20253
5.6 2.0287
};
\addplot [ultra thick, steelblue76114176]
table {%
6 2.0287
6 1.87854
};
\addplot [ultra thick, steelblue76114176]
table {%
6 2.20253
6 2.30966
};
\addplot [ultra thick, steelblue76114176]
table {%
5.8 1.87854
6.2 1.87854
};
\addplot [ultra thick, steelblue76114176]
table {%
5.8 2.30966
6.2 2.30966
};
\addplot [black, mark=o, mark size=3, mark options={solid,fill opacity=0,draw=steelblue76114176}, only marks]
table {%
6 2.5742
6 2.55496
};
\addplot [ultra thick, steelblue76114176]
table {%
-0.4 2.90149
0.4 2.90149
};
\addplot [forestgreen4416044, mark=triangle*, mark size=3, mark options={solid}, only marks]
table {%
0 2.92690892857143
};
\addplot [ultra thick, steelblue76114176]
table {%
0.6 2.980475
1.4 2.980475
};
\addplot [forestgreen4416044, mark=triangle*, mark size=3, mark options={solid}, only marks]
table {%
1 2.99821142857143
};
\addplot [ultra thick, indianred1967882]
table {%
1.6 3.099555
2.4 3.099555
};
\addplot [forestgreen4416044, mark=triangle*, mark size=3, mark options={solid}, only marks]
table {%
2 3.11136428571429
};
\addplot [ultra thick, indianred1967882]
table {%
2.6 2.90971
3.4 2.90971
};
\addplot [forestgreen4416044, mark=triangle*, mark size=3, mark options={solid}, only marks]
table {%
3 2.88819357142857
};
\addplot [ultra thick, indianred1967882]
table {%
3.6 3.00498
4.4 3.00498
};
\addplot [forestgreen4416044, mark=triangle*, mark size=3, mark options={solid}, only marks]
table {%
4 3.00703428571429
};
\addplot [ultra thick, steelblue76114176]
table {%
4.6 2.43929
5.4 2.43929
};
\addplot [forestgreen4416044, mark=triangle*, mark size=3, mark options={solid}, only marks]
table {%
5 2.44531785714286
};
\addplot [ultra thick, steelblue76114176]
table {%
5.6 2.11327
6.4 2.11327
};
\addplot [forestgreen4416044, mark=triangle*, mark size=3, mark options={solid}, only marks]
table {%
6 2.12731071428571
};
\end{axis}

\end{tikzpicture}

%% file: tikz/exp3_visqol.tex
% This file was created with tikzplotlib v0.10.1.
\begin{tikzpicture}

\definecolor{darkgray141160203}{RGB}{141,160,203}
\definecolor{darkgray176}{RGB}{176,176,176}
\definecolor{forestgreen4416044}{RGB}{44,160,44}
\definecolor{lightgray204}{RGB}{204,204,204}
\definecolor{mediumaquamarine102194165}{RGB}{102,194,165}
\definecolor{salmon25214198}{RGB}{252,141,98}

\begin{axis}[
tick align=outside,
tick pos=left,
x grid style={darkgray176},
xmajorgrids,
xmin=-0.5, xmax=3.5,
xtick style={color=black},
xticklabel style={rotate=90.0, font=\plotTickFont},
xticklabels={, \mTm{}, \shortstack{\mTm{}\\fine-tuned}, retr. \EC{}, \QGAN{}},
y grid style={darkgray176},
ymajorgrids,
ylabel={},
yticklabels={},
ytick style={color=black},
ymin=1.72, ymax=4.0,
]

\addplot [ultra thick, mediumaquamarine102194165, forget plot]
table {%
-0.4 2.336335
-0.133333333333333 2.336335
-0.133333333333333 2.49289
-0.4 2.49289
-0.4 2.336335
};
\addplot [ultra thick, mediumaquamarine102194165, forget plot]
table {%
-0.266666666666667 2.336335
-0.266666666666667 2.21777
};
\addplot [ultra thick, mediumaquamarine102194165, forget plot]
table {%
-0.266666666666667 2.49289
-0.266666666666667 2.62392
};
\addplot [ultra thick, mediumaquamarine102194165, forget plot]
table {%
-0.333333333333333 2.21777
-0.2 2.21777
};
\addplot [ultra thick, mediumaquamarine102194165, forget plot]
table {%
-0.333333333333333 2.62392
-0.2 2.62392
};
\addplot [black, mark=o, mark size=3, mark options={solid,fill opacity=0,draw=mediumaquamarine102194165}, only marks, forget plot]
table {%
-0.266666666666667 2.06083
};
\addplot [ultra thick, mediumaquamarine102194165, forget plot]
table {%
0.6 2.414745
0.866666666666667 2.414745
0.866666666666667 2.57244
0.6 2.57244
0.6 2.414745
};
\addplot [ultra thick, mediumaquamarine102194165, forget plot]
table {%
0.733333333333333 2.414745
0.733333333333333 2.26217
};
\addplot [ultra thick, mediumaquamarine102194165, forget plot]
table {%
0.733333333333333 2.57244
0.733333333333333 2.75643
};
\addplot [ultra thick, mediumaquamarine102194165, forget plot]
table {%
0.666666666666667 2.26217
0.8 2.26217
};
\addplot [ultra thick, mediumaquamarine102194165, forget plot]
table {%
0.666666666666667 2.75643
0.8 2.75643
};
\addplot [black, mark=o, mark size=3, mark options={solid,fill opacity=0,draw=mediumaquamarine102194165}, only marks, forget plot]
table {%
0.733333333333333 2.84595
};
\addplot [ultra thick, mediumaquamarine102194165, forget plot]
table {%
1.6 2.7697975
1.86666666666667 2.7697975
1.86666666666667 3.02731
1.6 3.02731
1.6 2.7697975
};
\addplot [ultra thick, mediumaquamarine102194165, forget plot]
table {%
1.73333333333333 2.7697975
1.73333333333333 2.69673
};
\addplot [ultra thick, mediumaquamarine102194165, forget plot]
table {%
1.73333333333333 3.02731
1.73333333333333 3.34671
};
\addplot [ultra thick, mediumaquamarine102194165, forget plot]
table {%
1.66666666666667 2.69673
1.8 2.69673
};
\addplot [ultra thick, mediumaquamarine102194165, forget plot]
table {%
1.66666666666667 3.34671
1.8 3.34671
};
\addplot [ultra thick, mediumaquamarine102194165, forget plot]
table {%
2.6 2.34539
2.86666666666667 2.34539
2.86666666666667 2.569845
2.6 2.569845
2.6 2.34539
};
\addplot [ultra thick, mediumaquamarine102194165, forget plot]
table {%
2.73333333333333 2.34539
2.73333333333333 2.1574
};
\addplot [ultra thick, mediumaquamarine102194165, forget plot]
table {%
2.73333333333333 2.569845
2.73333333333333 2.6543
};
\addplot [ultra thick, mediumaquamarine102194165, forget plot]
table {%
2.66666666666667 2.1574
2.8 2.1574
};
\addplot [ultra thick, mediumaquamarine102194165, forget plot]
table {%
2.66666666666667 2.6543
2.8 2.6543
};
\addplot [ultra thick, salmon25214198, forget plot]
table {%
-0.133333333333333 2.82464
0.133333333333333 2.82464
0.133333333333333 3.0300575
-0.133333333333333 3.0300575
-0.133333333333333 2.82464
};
\addplot [ultra thick, salmon25214198, forget plot]
table {%
0 2.82464
0 2.54078
};
\addplot [ultra thick, salmon25214198, forget plot]
table {%
0 3.0300575
0 3.20434
};
\addplot [ultra thick, salmon25214198, forget plot]
table {%
-0.0666666666666667 2.54078
0.0666666666666667 2.54078
};
\addplot [ultra thick, salmon25214198, forget plot]
table {%
-0.0666666666666667 3.20434
0.0666666666666667 3.20434
};
\addplot [ultra thick, salmon25214198, forget plot]
table {%
0.866666666666667 2.9338625
1.13333333333333 2.9338625
1.13333333333333 3.0295975
0.866666666666667 3.0295975
0.866666666666667 2.9338625
};
\addplot [ultra thick, salmon25214198, forget plot]
table {%
1 2.9338625
1 2.85716
};
\addplot [ultra thick, salmon25214198, forget plot]
table {%
1 3.0295975
1 3.15033
};
\addplot [ultra thick, salmon25214198, forget plot]
table {%
0.933333333333333 2.85716
1.06666666666667 2.85716
};
\addplot [ultra thick, salmon25214198, forget plot]
table {%
0.933333333333333 3.15033
1.06666666666667 3.15033
};
\addplot [black, mark=o, mark size=3, mark options={solid,fill opacity=0,draw=salmon25214198}, only marks, forget plot]
table {%
1 2.76701
1 3.37257
1 3.25144
1 3.21641
};
\addplot [ultra thick, salmon25214198, forget plot]
table {%
1.86666666666667 2.9709375
2.13333333333333 2.9709375
2.13333333333333 3.1904525
1.86666666666667 3.1904525
1.86666666666667 2.9709375
};
\addplot [ultra thick, salmon25214198, forget plot]
table {%
2 2.9709375
2 2.85418
};
\addplot [ultra thick, salmon25214198, forget plot]
table {%
2 3.1904525
2 3.3076
};
\addplot [ultra thick, salmon25214198, forget plot]
table {%
1.93333333333333 2.85418
2.06666666666667 2.85418
};
\addplot [ultra thick, salmon25214198, forget plot]
table {%
1.93333333333333 3.3076
2.06666666666667 3.3076
};
\addplot [black, mark=o, mark size=3, mark options={solid,fill opacity=0,draw=salmon25214198}, only marks, forget plot]
table {%
2 3.69904
};
\addplot [ultra thick, salmon25214198, forget plot]
table {%
2.86666666666667 2.78094
3.13333333333333 2.78094
3.13333333333333 2.9932775
2.86666666666667 2.9932775
2.86666666666667 2.78094
};
\addplot [ultra thick, salmon25214198, forget plot]
table {%
3 2.78094
3 2.57863
};
\addplot [ultra thick, salmon25214198, forget plot]
table {%
3 2.9932775
3 3.20405
};
\addplot [ultra thick, salmon25214198, forget plot]
table {%
2.93333333333333 2.57863
3.06666666666667 2.57863
};
\addplot [ultra thick, salmon25214198, forget plot]
table {%
2.93333333333333 3.20405
3.06666666666667 3.20405
};
\addplot [ultra thick, darkgray141160203, forget plot]
table {%
0.133333333333333 3.165975
0.4 3.165975
0.4 3.398135
0.133333333333333 3.398135
0.133333333333333 3.165975
};
\addplot [ultra thick, darkgray141160203, forget plot]
table {%
0.266666666666667 3.165975
0.266666666666667 2.994
};
\addplot [ultra thick, darkgray141160203, forget plot]
table {%
0.266666666666667 3.398135
0.266666666666667 3.60274
};
\addplot [ultra thick, darkgray141160203, forget plot]
table {%
0.2 2.994
0.333333333333333 2.994
};
\addplot [ultra thick, darkgray141160203, forget plot]
table {%
0.2 3.60274
0.333333333333333 3.60274
};
\addplot [ultra thick, darkgray141160203, forget plot]
table {%
1.13333333333333 3.293885
1.4 3.293885
1.4 3.4937775
1.13333333333333 3.4937775
1.13333333333333 3.293885
};
\addplot [ultra thick, darkgray141160203, forget plot]
table {%
1.26666666666667 3.293885
1.26666666666667 3.09354
};
\addplot [ultra thick, darkgray141160203, forget plot]
table {%
1.26666666666667 3.4937775
1.26666666666667 3.61701
};
\addplot [ultra thick, darkgray141160203, forget plot]
table {%
1.2 3.09354
1.33333333333333 3.09354
};
\addplot [ultra thick, darkgray141160203, forget plot]
table {%
1.2 3.61701
1.33333333333333 3.61701
};
\addplot [ultra thick, darkgray141160203, forget plot]
table {%
2.13333333333333 3.227385
2.4 3.227385
2.4 3.4498675
2.13333333333333 3.4498675
2.13333333333333 3.227385
};
\addplot [ultra thick, darkgray141160203, forget plot]
table {%
2.26666666666667 3.227385
2.26666666666667 2.92489
};
\addplot [ultra thick, darkgray141160203, forget plot]
table {%
2.26666666666667 3.4498675
2.26666666666667 3.61956
};
\addplot [ultra thick, darkgray141160203, forget plot]
table {%
2.2 2.92489
2.33333333333333 2.92489
};
\addplot [ultra thick, darkgray141160203, forget plot]
table {%
2.2 3.61956
2.33333333333333 3.61956
};
\addplot [black, mark=o, mark size=3, mark options={solid,fill opacity=0,draw=darkgray141160203}, only marks, forget plot]
table {%
2.26666666666667 3.85596
};
\addplot [ultra thick, darkgray141160203, forget plot]
table {%
3.13333333333333 3.0859575
3.4 3.0859575
3.4 3.2793675
3.13333333333333 3.2793675
3.13333333333333 3.0859575
};
\addplot [ultra thick, darkgray141160203, forget plot]
table {%
3.26666666666667 3.0859575
3.26666666666667 2.89736
};
\addplot [ultra thick, darkgray141160203, forget plot]
table {%
3.26666666666667 3.2793675
3.26666666666667 3.52747
};
\addplot [ultra thick, darkgray141160203, forget plot]
table {%
3.2 2.89736
3.33333333333333 2.89736
};
\addplot [ultra thick, darkgray141160203, forget plot]
table {%
3.2 3.52747
3.33333333333333 3.52747
};
\addplot [ultra thick, mediumaquamarine102194165, forget plot]
table {%
-0.4 2.41895
-0.133333333333333 2.41895
};
\addplot [forestgreen4416044, mark=triangle*, mark size=3, mark options={solid}, only marks, forget plot]
table {%
-0.266666666666667 2.40804107142857
};
\addplot [ultra thick, mediumaquamarine102194165, forget plot]
table {%
0.6 2.519465
0.866666666666667 2.519465
};
\addplot [forestgreen4416044, mark=triangle*, mark size=3, mark options={solid}, only marks, forget plot]
table {%
0.733333333333333 2.5125925
};
\addplot [ultra thick, mediumaquamarine102194165, forget plot]
table {%
1.6 2.857445
1.86666666666667 2.857445
};
\addplot [forestgreen4416044, mark=triangle*, mark size=3, mark options={solid}, only marks, forget plot]
table {%
1.73333333333333 2.89962035714286
};
\addplot [ultra thick, mediumaquamarine102194165, forget plot]
table {%
2.6 2.425255
2.86666666666667 2.425255
};
\addplot [forestgreen4416044, mark=triangle*, mark size=3, mark options={solid}, only marks, forget plot]
table {%
2.73333333333333 2.43766214285714
};
\addplot [ultra thick, salmon25214198, forget plot]
table {%
-0.133333333333333 2.90149
0.133333333333333 2.90149
};
\addplot [forestgreen4416044, mark=triangle*, mark size=3, mark options={solid}, only marks, forget plot]
table {%
0 2.92690892857143
};
\addplot [ultra thick, salmon25214198, forget plot]
table {%
0.866666666666667 2.980475
1.13333333333333 2.980475
};
\addplot [forestgreen4416044, mark=triangle*, mark size=3, mark options={solid}, only marks, forget plot]
table {%
1 2.99821142857143
};
\addplot [ultra thick, salmon25214198, forget plot]
table {%
1.86666666666667 3.050835
2.13333333333333 3.050835
};
\addplot [forestgreen4416044, mark=triangle*, mark size=3, mark options={solid}, only marks, forget plot]
table {%
2 3.09839571428571
};
\addplot [ultra thick, salmon25214198, forget plot]
table {%
2.86666666666667 2.90971
3.13333333333333 2.90971
};
\addplot [forestgreen4416044, mark=triangle*, mark size=3, mark options={solid}, only marks, forget plot]
table {%
3 2.88819357142857
};
\addplot [ultra thick, darkgray141160203, forget plot]
table {%
0.133333333333333 3.31666
0.4 3.31666
};
\addplot [forestgreen4416044, mark=triangle*, mark size=3, mark options={solid}, only marks, forget plot]
table {%
0.266666666666667 3.29474964285714
};
\addplot [ultra thick, darkgray141160203, forget plot]
table {%
1.13333333333333 3.396005
1.4 3.396005
};
\addplot [forestgreen4416044, mark=triangle*, mark size=3, mark options={solid}, only marks, forget plot]
table {%
1.26666666666667 3.38659535714286
};
\addplot [ultra thick, darkgray141160203, forget plot]
table {%
2.13333333333333 3.31546
2.4 3.31546
};
\addplot [forestgreen4416044, mark=triangle*, mark size=3, mark options={solid}, only marks, forget plot]
table {%
2.26666666666667 3.33104035714286
};
\addplot [ultra thick, darkgray141160203, forget plot]
table {%
3.13333333333333 3.20132
3.4 3.20132
};
\addplot [forestgreen4416044, mark=triangle*, mark size=3, mark options={solid}, only marks, forget plot]
table {%
3.26666666666667 3.18668321428571
};
\end{axis}

\end{tikzpicture}

%% file: tikz/lt1.tex
% This file was created with tikzplotlib v0.10.1.
\begin{tikzpicture}

\definecolor{darkgray176}{RGB}{176,176,176}
\definecolor{darkseagreen}{RGB}{143,188,143}
\definecolor{darkviolet}{RGB}{148,0,211}
\definecolor{gray}{RGB}{128,128,128}
\definecolor{limegreen}{RGB}{50,205,50}
\definecolor{magenta}{RGB}{255,0,255}
\definecolor{orange}{RGB}{255,165,0}
\definecolor{violet}{RGB}{238,130,238}

\begin{axis}[
tick align=outside,
tick pos=left,
x grid style={darkgray176},
xmin=-0.0400000000000001, xmax=14.04,
xtick style={color=black},
xtick={1,3,4,5,7,8,9,11,12,13},
xticklabel style={rotate=45.0},
xticklabels={
  DIRECT,
  10dB,
  20dB,
  30dB,
  \mTw{},
  \mTm{},
  \mTl{},
  \lTw{},
  \lTm{},
  \lTl{}
},
y grid style={darkgray176},
ylabel={DMOS},
ylabel style={at={(0.08,0.5)}},
ymajorgrids,
ymin=0.5, ymax=5.5,
ytick style={color=black},
ytick={1,2,3,4,5},
yticklabels={
  1, 2, 3, 4, 5
 % Very annoying=1,
  %Annoying=2,
  %Slighlty annoying=3,
  %Audible but not annoying=4,
  %Inaudible=5
}
]
\draw[draw=none,fill=gray] (axis cs:0.6,0) rectangle (axis cs:1.4,4.38333333333333);
\draw[draw=none,fill=orange] (axis cs:2.6,0) rectangle (axis cs:3.4,1.61666666666667);
\draw[draw=none,fill=orange] (axis cs:3.6,0) rectangle (axis cs:4.4,3.45);
\draw[draw=none,fill=orange] (axis cs:4.6,0) rectangle (axis cs:5.4,4.21666666666667);

\node at (axis cs:4,1.1) [draw, fill=orange!50, anchor=north, font=\small] {MNRU};

\draw[draw=none,fill=darkseagreen] (axis cs:6.6,0) rectangle (axis cs:7.4,3.01666666666667);
\draw[draw=none,fill=limegreen] (axis cs:7.6,0) rectangle (axis cs:8.4,3.56666666666667);
\draw[draw=none,fill=green] (axis cs:8.6,0) rectangle (axis cs:9.4,2);
\draw[draw=none,fill=darkviolet] (axis cs:10.6,0) rectangle (axis cs:11.4,3.21666666666667);
\draw[draw=none,fill=violet] (axis cs:11.6,0) rectangle (axis cs:12.4,3.43333333333333);
\draw[draw=none,fill=magenta] (axis cs:12.6,0) rectangle (axis cs:13.4,2.75);
\path [draw=black, semithick]
(axis cs:1,4.22133426333066)
--(axis cs:1,4.54533240333601);

\path [draw=black, semithick]
(axis cs:3,1.42978360001097)
--(axis cs:3,1.80354973332236);

\path [draw=black, semithick]
(axis cs:4,3.21509085780935)
--(axis cs:4,3.68490914219065);

\path [draw=black, semithick]
(axis cs:5,4.01850998229017)
--(axis cs:5,4.41482335104317);

\path [draw=black, semithick]
(axis cs:7,2.75111108589619)
--(axis cs:7,3.28222224743714);

\path [draw=black, semithick]
(axis cs:8,3.35641985627326)
--(axis cs:8,3.77691347706008);

\path [draw=black, semithick]
(axis cs:9,1.74911861582303)
--(axis cs:9,2.25088138417697);

\path [draw=black, semithick]
(axis cs:11,2.96974557326769)
--(axis cs:11,3.46358776006565);

\path [draw=black, semithick]
(axis cs:12,3.23367630160104)
--(axis cs:12,3.63299036506563);

\path [draw=black, semithick]
(axis cs:13,2.50513845720236)
--(axis cs:13,2.99486154279764);

\draw (axis cs:1,4.38333333333333) ++(0pt,16pt) node[
  scale=0.5,
  anchor=base,
  text=black,
  rotate=0.0
]{4.38};
\draw (axis cs:3,1.61666666666667) ++(0pt,16pt) node[
  scale=0.5,
  anchor=base,
  text=black,
  rotate=0.0
]{1.62};
\draw (axis cs:4,3.45) ++(0pt,16pt) node[
  scale=0.5,
  anchor=base,
  text=black,
  rotate=0.0
]{3.45};
\draw (axis cs:5,4.21666666666667) ++(0pt,16pt) node[
  scale=0.5,
  anchor=base,
  text=black,
  rotate=0.0
]{4.22};
\draw (axis cs:7,3.01666666666667) ++(0pt,16pt) node[
  scale=0.5,
  anchor=base,
  text=black,
  rotate=0.0
]{3.02};
\draw (axis cs:8,3.56666666666667) ++(0pt,16pt) node[
  scale=0.5,
  anchor=base,
  text=black,
  rotate=0.0
]{3.57};
\draw (axis cs:9,2) ++(0pt,16pt) node[
  scale=0.5,
  anchor=base,
  text=black,
  rotate=0.0
]{2.00};
\draw (axis cs:11,3.21666666666667) ++(0pt,16pt) node[
  scale=0.5,
  anchor=base,
  text=black,
  rotate=0.0
]{3.22};
\draw (axis cs:12,3.43333333333333) ++(0pt,16pt) node[
  scale=0.5,
  anchor=base,
  text=black,
  rotate=0.0
]{3.43};
\draw (axis cs:13,2.75) ++(0pt,16pt) node[
  scale=0.5,
  anchor=base,
  text=black,
  rotate=0.0
]{2.75};
\end{axis}

\end{tikzpicture}

%% file: tikz/lt2.tex
% This file was created with tikzplotlib v0.10.1.
\begin{tikzpicture}

\definecolor{cyan}{RGB}{0,255,255}
\definecolor{darkgray176}{RGB}{176,176,176}
\definecolor{darkred}{RGB}{139,0,0}
\definecolor{gray}{RGB}{128,128,128}
\definecolor{magenta}{RGB}{255,0,255}
\definecolor{orange}{RGB}{255,165,0}
\definecolor{pink}{RGB}{255,192,203}
\definecolor{violet}{RGB}{238,130,238}

\newcommand{\whitePerc}{50}

\begin{axis}[
tick align=outside,
tick pos=left,
x grid style={darkgray176},
xmin=-0.64, xmax=26.64,
xtick style={color=black},
xtick={1,3,4,5,7,8,19,20,21,23,24,25,9,11,12,13,15,16,17},
xticklabel style={rotate=90.0, font=\small},
xticklabels={
  DIRECT,
  10dB,
  20dB,
  30dB,
  \MBD{},
  \EC{},
  1.5 kbps,
  3 kbps,
  6 kbps,
  1.5 kbps,
  3 kbps,
  6 kbps,
  \LDC{},
  1.5 kbps,
  3 kbps,
  6 kbps,
  1.5 kbps,
  3 kbps,
  6 kbps
},
y grid style={darkgray176},
ylabel={DMOS},
ylabel style={at={(0.08,0.5)}},
ymajorgrids,
ymin=0.5, ymax=5.0,
ytick style={color=black},
ytick={1,2,3,4,5},
yticklabels={
  1, 2, 3, 4, 5
 % Very annoying=1,
  %Annoying=2,
  %Slighlty annoying=3,
  %Audible but not annoying=4,
  %Inaudible=5
}
]

\draw[draw=none,fill=gray] (axis cs:0.6,0) rectangle (axis cs:1.4,4.27631578947368);

\draw[draw=none,fill=orange] (axis cs:2.6,0) rectangle (axis cs:3.4,1.39473684210526);
\draw[draw=none,fill=orange] (axis cs:3.6,0) rectangle (axis cs:4.4,3.06578947368421);
\draw[draw=none,fill=orange] (axis cs:4.6,0) rectangle (axis cs:5.4,4.06578947368421);

\node at (axis cs:4,1.1) [draw, fill=orange!\whitePerc, anchor=north, font=\tiny] {MNRU};

\draw[draw=none,fill=darkred] (axis cs:6.6,0) rectangle (axis cs:7.4,1.36842105263158);
% was magenta
\draw[draw=none,fill=darkred] (axis cs:7.6,0) rectangle (axis cs:8.4,1.51315789473684);
% was cyan
\draw[draw=none,fill=darkred] (axis cs:8.6,0) rectangle (axis cs:9.4,1.85526315789474);

\node at (axis cs:8,1.1) [draw, fill=darkred!\whitePerc, anchor=north, font=\tiny, align=center] {pretrained\\3 kbps};

\draw[draw=none,fill=blue] (axis cs:18.6,0) rectangle (axis cs:19.4,2.68421052631579);
\draw[draw=none,fill=blue] (axis cs:19.6,0) rectangle (axis cs:20.4,3.11842105263158);
\draw[draw=none,fill=blue] (axis cs:20.6,0) rectangle (axis cs:21.4,3.13157894736842);

\node at (axis cs:20,1.1) [draw, fill=blue!\whitePerc, anchor=north, font={\normalfont\tiny}] {\mTm{}};

\draw[draw=none,fill=violet] (axis cs:22.6,0) rectangle (axis cs:23.4,3.05263157894737);
\draw[draw=none,fill=violet] (axis cs:23.6,0) rectangle (axis cs:24.4,3.21052631578947);
\draw[draw=none,fill=violet] (axis cs:24.6,0) rectangle (axis cs:25.4,3.34210526315789);

\node at (axis cs:24,1.1) [draw, fill=violet!\whitePerc, anchor=north, font={\normalfont\tiny}, align=center] {\mTm{}\\fine-tuned};

\draw[draw=none,fill=pink] (axis cs:10.6,0) rectangle (axis cs:11.4,2.64473684210526);
\draw[draw=none,fill=pink] (axis cs:11.6,0) rectangle (axis cs:12.4,3.06578947368421);
\draw[draw=none,fill=pink] (axis cs:12.6,0) rectangle (axis cs:13.4,3.18421052631579);

\node at (axis cs:12,1.1) [draw, fill=pink!\whitePerc, anchor=north, font={\normalfont\tiny}] {\QGAN{}};

\draw[draw=none,fill=gray] (axis cs:14.6,0) rectangle (axis cs:15.4,3.53947368421053);
\draw[draw=none,fill=gray] (axis cs:15.6,0) rectangle (axis cs:16.4,3.55263157894737);
\draw[draw=none,fill=gray] (axis cs:16.6,0) rectangle (axis cs:17.4,3.77631578947368);

\node at (axis cs:16,1.1) [draw, fill=gray!\whitePerc, anchor=north, font={\normalfont\tiny}, align=center] {\EC{}\\retrained};

\path [draw=black, semithick]
(axis cs:1,4.10178372811184)
--(axis cs:1,4.45084785083552);

\path [draw=black, semithick]
(axis cs:3,1.26196070256294)
--(axis cs:3,1.52751298164759);

\path [draw=black, semithick]
(axis cs:4,2.82410114466496)
--(axis cs:4,3.30747780270346);

\path [draw=black, semithick]
(axis cs:5,3.87035585064125)
--(axis cs:5,4.26122309672717);

\path [draw=black, semithick]
(axis cs:7,1.20080931291488)
--(axis cs:7,1.53603279234828);

\path [draw=black, semithick]
(axis cs:8,1.35105988851969)
--(axis cs:8,1.67525590095399);

\path [draw=black, semithick]
(axis cs:19,2.46328237422983)
--(axis cs:19,2.90513867840175);

\path [draw=black, semithick]
(axis cs:20,2.880579271536)
--(axis cs:20,3.35626283372716);

\path [draw=black, semithick]
(axis cs:21,2.88709981499093)
--(axis cs:21,3.37605807974591);

\path [draw=black, semithick]
(axis cs:23,2.82962593740703)
--(axis cs:23,3.27563722048771);

\path [draw=black, semithick]
(axis cs:24,2.99861278092115)
--(axis cs:24,3.4224398506578);

\path [draw=black, semithick]
(axis cs:25,3.1201359358366)
--(axis cs:25,3.56407459047919);

\path [draw=black, semithick]
(axis cs:9,1.64853943836981)
--(axis cs:9,2.06198687741966);

\path [draw=black, semithick]
(axis cs:11,2.39383119533214)
--(axis cs:11,2.89564248887839);

\path [draw=black, semithick]
(axis cs:12,2.83261687900286)
--(axis cs:12,3.29896206836556);

\path [draw=black, semithick]
(axis cs:13,2.97104527628025)
--(axis cs:13,3.39737577635133);

\path [draw=black, semithick]
(axis cs:15,3.31482383046763)
--(axis cs:15,3.76412353795342);

\path [draw=black, semithick]
(axis cs:16,3.35699387130887)
--(axis cs:16,3.74826928658587);

\path [draw=black, semithick]
(axis cs:17,3.57997674753199)
--(axis cs:17,3.97265483141538);

\draw (axis cs:1,4.27631578947368) ++(0pt,16pt) node[
  scale=0.5,
  anchor=base,
  text=black,
  rotate=60.0
]{4.28};
\draw (axis cs:3,1.39473684210526) ++(0pt,16pt) node[
  scale=0.5,
  anchor=base,
  text=black,
  rotate=60.0
]{1.39};
\draw (axis cs:4,3.06578947368421) ++(0pt,16pt) node[
  scale=0.5,
  anchor=base,
  text=black,
  rotate=60.0
]{3.07};
\draw (axis cs:5,4.06578947368421) ++(0pt,16pt) node[
  scale=0.5,
  anchor=base,
  text=black,
  rotate=60.0
]{4.07};
\draw (axis cs:7,1.36842105263158) ++(0pt,16pt) node[
  scale=0.5,
  anchor=base,
  text=black,
  rotate=60.0
]{1.37};
\draw (axis cs:8,1.51315789473684) ++(0pt,16pt) node[
  scale=0.5,
  anchor=base,
  text=black,
  rotate=60.0
]{1.51};
\draw (axis cs:19,2.68421052631579) ++(0pt,16pt) node[
  scale=0.5,
  anchor=base,
  text=black,
  rotate=60.0
]{2.68};
\draw (axis cs:20,3.11842105263158) ++(0pt,16pt) node[
  scale=0.5,
  anchor=base,
  text=black,
  rotate=60.0
]{3.12};
\draw (axis cs:21,3.13157894736842) ++(0pt,16pt) node[
  scale=0.5,
  anchor=base,
  text=black,
  rotate=60.0
]{3.13};
\draw (axis cs:23,3.05263157894737) ++(0pt,16pt) node[
  scale=0.5,
  anchor=base,
  text=black,
  rotate=60.0
]{3.05};
\draw (axis cs:24,3.21052631578947) ++(0pt,16pt) node[
  scale=0.5,
  anchor=base,
  text=black,
  rotate=60.0
]{3.21};
\draw (axis cs:25,3.34210526315789) ++(0pt,16pt) node[
  scale=0.5,
  anchor=base,
  text=black,
  rotate=60.0
]{3.34};
\draw (axis cs:9,1.85526315789474) ++(0pt,16pt) node[
  scale=0.5,
  anchor=base,
  text=black,
  rotate=60.0
]{1.86};
\draw (axis cs:11,2.64473684210526) ++(0pt,16pt) node[
  scale=0.5,
  anchor=base,
  text=black,
  rotate=60.0
]{2.64};
\draw (axis cs:12,3.06578947368421) ++(0pt,16pt) node[
  scale=0.5,
  anchor=base,
  text=black,
  rotate=60.0
]{3.07};
\draw (axis cs:13,3.18421052631579) ++(0pt,16pt) node[
  scale=0.5,
  anchor=base,
  text=black,
  rotate=60.0
]{3.18};
\draw (axis cs:15,3.53947368421053) ++(0pt,16pt) node[
  scale=0.5,
  anchor=base,
  text=black,
  rotate=60.0
]{3.54};
\draw (axis cs:16,3.55263157894737) ++(0pt,16pt) node[
  scale=0.5,
  anchor=base,
  text=black,
  rotate=60.0
]{3.55};
\draw (axis cs:17,3.77631578947368) ++(0pt,16pt) node[
  scale=0.5,
  anchor=base,
  text=black,
  rotate=60.0
]{3.78};
\end{axis}

\end{tikzpicture}